%% file: main.tex
\documentclass[sigconf, nonacm]{acmart}
\PassOptionsToPackage{hyphens}{url}
\usepackage{hyperref}
\usepackage{listings}
\usepackage{xcolor}
\usepackage[dvipsnames]{xcolor}
\usepackage{annotate-equations}
\usepackage{enumitem}
\usepackage{subcaption}
\usepackage{framed}

\definecolor{rqblue}{RGB}{17,147,180}   %
\definecolor{rqgray}{RGB}{242,242,242}  %

\newenvironment{rqbox}
  {
    \setlength{\fboxsep}{8pt}%
    
    \MakeFramed{%
      \advance\hsize-4pt%
      \advance\hsize-2\fboxsep%
      \FrameRestore
    }
    \noindent\ignorespaces
  }
  {
    \endMakeFramed
  }

\definecolor{kw}{RGB}{0,102,153}      %
\definecolor{str}{RGB}{196,  26, 22}  %
\definecolor{com}{RGB}{95,  95, 95}   %
\definecolor{ln}{RGB}{120,120,120}    %
\definecolor{rule}{RGB}{200,200,200}  %

\lstdefinestyle{acmcode}{
  basicstyle=\ttfamily\footnotesize,
  numbers=left,
  numberstyle=\tiny\color{ln},
  numbersep=6pt,
  keywordstyle=\bfseries\color{kw},
  commentstyle=\itshape\color{com},
  stringstyle=\color{str},
  showstringspaces=false,
  breaklines=true,
  breakatwhitespace=true,
  tabsize=2,
  columns=flexible,
  frame=lines,
  rulecolor=\color{rule},
  aboveskip=4pt,
  belowskip=4pt,
}

\newif\ifshowacm
\showacmfalse %

\ifshowacm
\else
  \renewcommand\footnotetextcopyrightpermission[1]{} %
  \renewcommand{\keywords}[1]{} %
\fi

\newlist{rqlist}{enumerate}{1}
\setlist[rqlist]{label*=\textbf{RQ\arabic*}}

\newcommand{\spackagent}[1]{SpackIt}

\begin{document}

\title{LLMs as Packagers of HPC Software}

\author{%
Caetano Melone, %
Daniel Nichols, %
Konstantinos Parasyris, %
Todd Gamblin, %
Harshitha Menon%
\\[0.4em]
{\fontsize{10pt}{11pt}\selectfont\emph{Lawrence Livermore National Laboratory}}%
\\[0.4em]
{\fontsize{10pt}{11pt}\selectfont
Email:
\href{mailto:cmelone@llnl.gov}{cmelone@llnl.gov},\,
\href{mailto:harshitha@llnl.gov}{harshitha@llnl.gov}}%
\\[0.9em] 
} %

\renewcommand{\shortauthors}{Melone et al.}

\begin{abstract}
\input{abstract} 
\end{abstract}

\maketitle

\section{Introduction}
\label{sec:intro}
\input{intro}

\section{Background}
\label{sec:background}
\input{background}

\section{Approaches for Package Generation}
\label{sec:approaches}
\input{approaches}

\section{Metrics for Evaluation}
\label{sec:metrics}
\input{metrics}

\section{Set of Spack Package Generation Tasks}
\label{sec:tasks}
\input{tasks}

\section{Experimental Setup}
\label{sec:setup}
\input{setup}

\section{Results}
\label{sec:results}
\input{results}

\section{Threats to Validity}
\label{sec:threats}
\input{threats}

\section{Related Work}
\label{sec:related}
\input{related}

\section{Conclusion}
\label{sec:conclusion}
\input{conclusion}

\begin{acks}
\input{acknowledgements}
\end{acks}
\bibliographystyle{ACM-Reference-Format}
\bibliography{bib}
\clearpage
\appendix
\input{appendix}

\end{document}

%% file: abstract.tex
High performance computing (HPC) software ecosystems are inherently heterogeneous, comprising scientific applications that depend on hundreds of external packages, each with distinct build systems, options, and dependency constraints. Tools such as Spack automate dependency resolution and environment management, but their effectiveness relies on manually written build recipes. As these ecosystems grow, maintaining existing specifications and creating new ones becomes increasingly labor-intensive. While large language models (LLMs) have shown promise in code generation, automatically producing correct and maintainable Spack recipes remains a significant challenge. We present a systematic analysis of how LLMs and context-augmentation methods can assist in the generation of Spack recipes. To this end, we introduce \spackagent{}, an end-to-end framework that combines repository analysis, retrieval of relevant examples, and iterative refinement through diagnostic feedback. We apply \spackagent{} to a representative subset of 308 open-source HPC packages to assess its effectiveness and limitations. Our results show that \spackagent{} increases installation success from 20\% in a zero-shot setting to over 80\% in its best configuration, demonstrating the value of retrieval and structured feedback for reliable package synthesis.

%% file: intro.tex
Software engineering has long emphasized constructing systems from reusable components~\cite{mcilroy1968mass}. Component reuse accelerates development, improves maintainability, and promotes separation of concerns.

However, this efficiency of reuse has also introduced substantial complexity into software ecosystems~\cite{lehman1980programs, gonzalez2009macro, decan2019empirical, bogart2016break, dietrich2019dependency}.
Package managers (e.g., APT, Conda, Cargo, npm, and Spack) have become critical to managing this complexity, enabling large-scale software reuse with millions of components, commonly referred to as packages, through build specifications.

These specifications are typically expressed in structured metadata files (e.g., \emph{package.json} in npm) or through domain-specific languages, such as Spack’s~\cite{gamblin_spack_2015} \emph{recipes}, which describe dependency graphs and configuration variants. While such representations enable automation and reproducibility, crafting detailed metadata requires a deep understanding of the project's build system, dependency graph, and platform constraints, as well as familiarity with each package manager's conventions for expressing these details.

This challenge is particularly acute in high performance computing (HPC), where software 

comprises hundreds of interdependent components with platform-specific build options, compiler constraints, and hardware-dependent optimizations. Consequently, package engineering in HPC demands a high degree of configurability and domain expertise, making the creation and maintenance of software packages a complex and error-prone process. Moreover, package creation is not a one-shot task. It requires evaluation, feedback, and refinement to ensure correctness and portability across diverse architectures.

The significant effort required to create and maintain packages motivates exploration of automated approaches.

Recent advances in large language models (LLMs) have transformed software development workflows, supporting tasks such as code completion, repository analysis, and translation. 
As LLMs are adapted for domain-specific applications, retrieval-augmented generation (RAG) has emerged as a technique to improve factual accuracy and domain relevance~\cite{lewis_retrieval-augmented_2020, guo_lightrag_2025}. RAG methods integrate structured domain knowledge into a model’s context, which is defined as the information provided as input to the model, to better guide reasoning and generation. This additional context helps reduce hallucination~\cite{spracklen_we_2025} and improves performance on specialized tasks beyond the model’s native capability. This presents an exciting opportunity: the potential to leverage variations of these techniques to automatically generate valid package installation recipes for the HPC ecosystem.
However, while prior work has explored LLM-assisted code generation in open-source contexts, little is known about their effectiveness within package management ecosystems, particularly in fields like HPC and scientific computing, where domain-specific code, metadata, and build logic are largely limited from public training corpora.

In this work, we investigate the use of large language models for automating package specification generation within the HPC ecosystem. We present \spackagent{}, a tool-augmented LLM agent designed to synthesize valid Spack package recipes by emulating the workflow of an expert packager. The agent decomposes the task into interpretable steps: inspecting repositories to collect build metadata and directory structures, retrieving similar examples from a curated knowledge base, drafting candidate recipes, and iteratively refining them through build-and-debug cycles guided by structured feedback. To assess the impact of augmentation strategies such as RAG and self-repair loops, we conduct a large-scale empirical study on a representative subset of packages. Our evaluation measures the syntactic, semantic, and build correctness of generated recipes across multiple LLMs and context-enhancement configurations. Through this study, we provide new insights into how retrieval and agentic workflows influence LLM performance in complex software engineering domains. 

Our main contributions are as follows:%
\begin{itemize}[leftmargin=*]
\item A novel tool-augmented agent for automating Spack package recipe generation.
\item An agent architecture that integrates repository analysis, metadata distillation, and multiple retrieval strategies to provide relevant context and ground model reasoning during generation.
\item Use of a self-repair loop that uses feedback to iteratively refine incorrect recipes. 
\item A large-scale empirical study of LLM performance on 308 scientific software packages from the E4S stack, evaluating multiple models and context-enhancement strategies.
\item Evaluation metrics designed to quantify the semantic and functional correctness of generated packages.
\end{itemize}

\newcommand{\RQone}{How can we best distill large, complex repository metadata to enable language models to reason about and generate structured build specifications for scientific software?}

\newcommand{\RQtwo}{How well can language models fix errors in their own generated package recipes?}

\newcommand{\RQthree}{Does a cleaner error signal improve velocity in model refinement?}

\newcommand{\RQfour}{Can language models with context augmentation and iterative repair loops outperform state-of-the-art, general-purpose software engineering agents in Spack package recipe generation in terms of efficacy and cost?}

To systematically assess \spackagent{}’s capabilities and examine the effects of retrieval and repair strategies, we pose the following research questions:
\begin{rqlist}
  \item \emph{\RQone}
  \item \emph{\RQtwo}
  \item \emph{\RQthree}
  \item \emph{\RQfour}
\end{rqlist}

\noindent\textbf{Replication Package}: Our code and data are shared at \url{https://github.com/spack/spack-it}.

%% file: background.tex
In this section, we present an overview of the Spack package manager and provide background on LLMs for software engineering.%
\subsection{Spack package manager}
\label{sec:bg-spack}
Modern software ecosystems rely on package managers to automate the installation and configuration of dependencies and build options. These systems ensure that complex software stacks can be reproduced consistently across environments by resolving dependency graphs, enforcing version constraints, and orchestrating build processes. 
Such capabilities are crucial for the development and execution of scientific applications, where reproducibility and portability across architectures and machines are critical.

In this paper, we use the Spack \cite{gamblin_spack_2015} package manager to model relationships between independent software components, execute installation logic, and evaluate whether automatically generated build specifications can be used to install software. Spack was designed to overcome the persistent challenges of building scientific software \cite{dubois_why_2003, hoste_easybuild_2012}, providing flexibility in the selection of compilers, system libraries, and configuration options. 

Spack manages a large open-source ecosystem with over 8,500 packages spanning parallel programming models, communication libraries, machine learning frameworks, and scientific applications. Its declarative syntax allows users to encode complex build relationships while maintaining portability across systems, which is critical for HPC.

\begin{figure}[h]
  \centering
  \begin{lstlisting}[style=acmcode, language=Python]
class Example(CMakePackage, ROCmPackage, CudaPackage):
    version("1.0")

    variant("openmp", default=False)
    variant("hip", default=False)
    variant("cuda", default=False)

    conflicts("+cuda", when="+hip")

    depends_on("c")
    depends_on("mpi@3")

    def cmake_args(self):
        return [self.define_from_variant("ENABLE_OPENMP", "openmp")]
  \end{lstlisting}
  \caption{Example of a Spack package defining a software build specification. Each recipe is implemented as a Python class that declares versions, build variants, and dependencies. This structure enables configurability and reproducibility across installation targets.}
  \label{fig:spack-example}
  \Description{Python code defining build options and dependencies for an example Spack package.}
  \vspace{-2em}
\end{figure}

Each package in Spack is defined through a \textbf{\emph{recipe}}, a Python class that specifies versions, dependencies, build variants, and installation logic, as shown in Figure~\ref{fig:spack-example}. These recipes use declarative \emph{directives} to encode configuration constraints that may vary across platforms, compiler toolchains, or accelerator architectures. The example package defines optional CUDA and HIP support, with a conflict preventing both from being enabled simultaneously. It also requires an MPI implementation supporting version~3 and conditionally enables OpenMP support when requested by the user.

This approach to package management provides significant flexibility: users can build the same packages with different compilers, MPI implementations, or optional features while preserving reproducibility through dependency resolution. Unlike conventional package managers that target a single environment or are solely focused on binary installations, Spack was designed for heterogeneous HPC systems, making it a natural choice for studying automated generation of complex build specifications.

Directives and installation logic are written by package recipe maintainers, who spend considerable time analyzing repository files and documentation to ensure that each recipe remains functional and updated. Although Spack provides documentation and extensive tooling \cite{noauthor_spack_2025} to assist this process, recipe creation and maintenance remains labor-intensive and a high burden for newer members of the community. In this work, we view this manual authoring process as an opportunity for automation: by extracting and reasoning over repository metadata, we investigate whether language models can generate valid Spack packages.

\noindent\textbf{The Extreme-scale Scientific Software Stack (E4S).} 
E4S is an open-source ecosystem that curates and integrates a wide range of high performance computing (HPC) software packages \cite{noauthor_extreme-scale_nodate, heroux_scalable_2024}. It encompasses parallel programming models, numerical libraries, I/O and data analysis tools, and visualization frameworks, all designed for large-scale systems.

E4S uses the Spack package manager to automate the building and testing of these components, ensuring compatibility, portability, and reproducibility across diverse HPC architectures. In this work, we use E4S both as a source of reference data for retrieval and reasoning experiments and as the basis of our experimental task set for evaluating automated package generation and repair.

\subsection{LLMs for Software Engineering Tasks}
\label{sec:bg-llm}
An LLM \emph{agent} is a system that uses a language model to reason and execute multi-step tasks by using external tools. \textbf{SWE-agent} is a state-of-the-art agentic framework for automated software repair \cite{yang_swe-agent_2024}. It operates within a closed feedback loop that plans, applies changes, and verifies results. The agent excels at retrieving and reasoning over relevant information from the software project it is tasked with modifying and can coordinate edits across multiple components within a repository.
While not designed specifically for Spack or build specification synthesis, SWE-agent provides a useful comparative point for evaluating how a general-purpose, code-editing agent performs on structured build metadata. This contrast helps contextualize the distinction between domain-aware package generation and general software repair workflows.

%% file: approaches.tex
\begin{figure*}[t]
  \centering
  \includegraphics[width=0.87\linewidth]{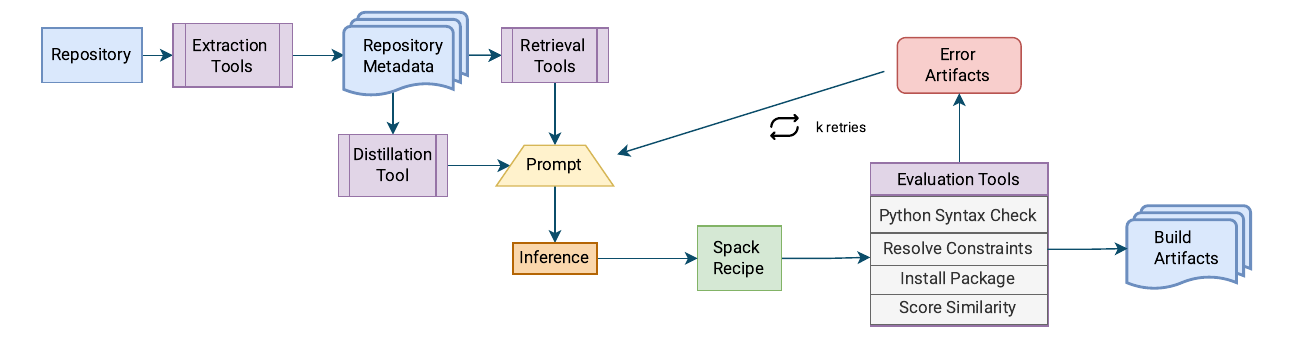}
  \caption{Overview of our approach. The pipeline is coordinated by a single agent that orchestrates specialized tools for extraction, retrieval, generation, evaluation, and repair. It begins by analyzing a repository to collect build metadata and directory structure. Examples from either graph-based or embedding-based retrieval tools are combined with this metadata to form the context for generation. The resulting recipe is then validated in an isolated build environment. If failures occur, the error-aware repair loop leverages feedback from Spack outputs to iteratively refine the package for up to $k$ attempts. }
  \label{fig:execution-flow}
\end{figure*}

Generating \textbf{Spack recipes}, defined in ~\ref{sec:bg-spack} as declarative build specifications that describe how to configure and compile software cannot be done in a single iteration. It requires a deep understanding of the project's build system, dependency graph, and platform constraints, as well as familiarity with Spack's conventions for expressing these details in a recipe.
Our goal with \spackagent{} is to generate valid \textbf{Spack recipes} and evaluate them in accordance with Spack's best-practices. To accomplish this, \spackagent{} is designed as an autonomous agent that leverages LLMs coupled with a suite of specialized tools. \spackagent{} receives a high-level goal, such as \emph{"Generate a valid Spack recipe for repository X"}, and decomposes the task into a sequence of retrieval, reasoning, and an iterative repair stages.
This agentic approach is model-agnostic, meaning it can operate with different LLMs. However, its overall performance and refinement ability improve with the reasoning and coding capabilities of the underlying LLM.
An overview of the resulting orchestration pipeline is presented in Figure~\ref{fig:execution-flow}. Additional configuration details are provided in Appendix~\ref{sec:config}.

Our agent's architecture is guided by three core principles:
\begin{itemize}[leftmargin=*]
    \item \textbf{Grounding in Source Repository:} An agent cannot reliably generate build recipes without extracting information from the repository. This requires  information extraction tools that inspect the repository, parse its content, and extract and distill important metadata from build files.
    \item \textbf{Leveraging Relevant Examples:} Developers rarely write Spack recipes from scratch. They look at examples to understand structure and best practices. In the same way, \spackagent{} benefits by using relevant references, such as existing recipes from comparable packages, to guide and improve its generation process.
    \item \textbf{Refinement via Iterative Feedback:} Generating valid Spack recipes is unlikely to be successful on the first attempt. It is important to be able to iteratively evaluate and refine the generated recipe. 
    \spackagent{} uses a self-repair loop that digests structured feedback from build results to iteratively refine recipes.
\end{itemize}

\subsection{Repository Metadata Extraction}

The first stage of our approach is orchestrated by a unified agent that coordinates specialized extraction tools. Given a target repository, the agent invokes these tools to identify and parse files capturing build and dependency logic, such as \texttt{CMakeLists.txt} and related configuration artifacts. It also maps the repository’s directory structure to preserve contextual relationships among source components. The resulting metadata encodes dependency declarations, compiler and platform constraints, and build features such as CUDA support or Python extensions. This information is then normalized into a compact, consistent schema, reducing input length while preserving the logical structure required for downstream heuristics.

Depending on the experimental configuration, an optional distillation tool further compresses the extracted metadata into concise summaries that capture relationships between configuration options and Spack concepts such as compiler flags, variants, and dependency constraints. This step is configurable, allowing experiments to compare runs with and without distillation to evaluate its impact on model performance. The resulting structured metadata serves as the foundation for subsequent stages of recipe synthesis.

\subsection{Retrieval and Context Augmentation}
\label{subsec:retrieval-tool}

The next stage of the pipeline is the retrieval stage, in which the agent uses tools to obtain relevant reference packages. Given evidence from the repository metadata, it queries relevant information to augment the context supplied to the LLM during inference. This design is motivated by the need to identify where LLMs struggle in package generation and to assess whether retrieving structurally similar recipes, together with repository metadata, can improve synthesis. For example, when generating a recipe for a physics simulation code with CUDA support, retrieving similar packages can reveal how to express  build directives that correctly specify CUDA toolchains and runtime libraries in Spack’s expected format. This grounding aligns the generated recipe with Spack conventions and exposes the model to current patterns for expressing comparable build constraints. Prior work finds that retrieval of similar examples improves code generation, yet retrievers often miss useful context and LLMs only partially exploit what is retrieved \cite{gu_what_2025, wang_coderag-bench_2025}, warranting further study of retrieval quality and integration in this setting. To isolate the marginal effects of different reference types, we enumerate several retrieval methods.

\textbf{Similar Reference Selection.} 
Spack packages encode information about their dependencies and supported features, such as platform-specific options and accelerator support. Referencing similar packages helps the LLM express comparable build constraints and variants in Spack’s expected format.

We draw on the E4S dataset of Spack packages described in~\ref{sec:bg-spack}, which provides a collection of widely used and actively maintained software recipes. These packages are subject to extensive community scrutiny and consistent testing, making them reliable reference points for our analysis. From this dataset, we extract metadata such as dependencies, constraints, recipes, and other build logic. The resulting information is stored in a Neo4j graph database where packages are represented as nodes and dependency relationships as edges. This forms a knowledge graph of Spack package metadata.

In order to select similar packages from the graph, we require a scalar value that quantifies how closely a candidate package resembles a given target. In the context of Spack packages, this ranking should (A) reflect structural similarity in dependency and build configurations, and (B) remain interpretable, enabling inspection of which components contribute to similarity. Thus, we define the \emph{Package Affinity Score}, a metric that ranks existing packages by how closely their dependency and build option structures align with those of the target.

The score is computed by comparing each candidate’s declared dependencies and
build options against those of the target and combining the resulting overlaps
through a weighted sum. Packages with the highest scores are considered the most
structurally similar and are selected as reference examples for subsequent
recipe generation. Target repositories are excluded from the search.
We formalize the metric as the \emph{Package Affinity Score}
$\mathcal{A}(p_{\text{t}}, p)$:

\begin{equation}
\label{eq:pkg-affinity-score}
\mathcal{A}(p_{\text{t}}, p)
= w_D \cdot \left| D_{\text{t}} \cap D_p \right|
+ w_B \cdot \left| B_{\text{t}} \cap B_p \right|,
\end{equation}

\noindent
where $D_{\text{t}}$ and $B_{\text{t}}$ denote the sets of dependencies and build options of the target repository, and $D_p$, $B_p$ are those of a candidate package $p$. The weights $w_D = 0.6$ and $w_B = 0.4$ balance the influence of each component, reflecting that dependencies govern the overall structure of a recipe, while build options capture user-level flexibility. The highest-scoring candidate serves as the reference example returned by the retrieval tool.

\noindent\textbf{Random Reference Selection} 
To test whether the effect of a similar recipe in the context is due to the fact that we included any Spack recipe, we support the inclusion of two types of random recipe selection. The retrieval tool can either receive any random recipe from our E4S knowledge graph, or a random recipe selected among packages of the same build system. For example, if the target repository was identified as supporting CMake, the random recipe selector will choose a package with that same build system.

\noindent\textbf{Embedding-Based Retrieval} 
In addition to the graph-based similarity method, we also evaluate a dense embedding-based retrieval approach using the same E4S dataset of Spack packages. Rather than computing similarity through explicit overlap in dependency or build metadata, this approach measures proximity between vector representations of package metadata and recipe text, capturing relationships that may not be reflected in structural features alone.

Our embedding-based method follows the canonical retrieve-then-generate paradigm commonly used in code generation settings \cite{wang_coderag-bench_2025,lewis_retrieval-augmented_2020}, often referred to as retrieval-augmented generation (RAG). 

To construct the retrieval corpus, each package is represented by two textual views. The first is a compact feature card summarizing the package's build system, dependencies, and build options. The second is a collection of recipe chunks obtained through structure-aware filtering of the Spack recipe source file. This segmentation identifies local regions such as variant declarations, dependency lists, and method overrides of installation or testing logic, ensuring that each chunk encodes a self-contained portion of build logic. Together, these representations can capture both declarative and procedural aspects of a package.

All feature cards and chunks are embedded using the Nomic Embed Code model, a 7-billion-parameter sentence-transformer for code and documentation embeddings \cite{suresh_cornstack_2025}, and stored in indices using cosine similarity. Each index is cached using an identifier derived from the dataset to ensure reproducibility. During retrieval, a target repository's metadata is represented as a query string describing dependencies, build systems, and other flags. The query embedding is compared to the stored vectors to rank similar packages, and the top-ranked recipes contribute representative chunks for context during generation. We use this model for its strong performance on code and documentation retrieval, making it a competitive open model for this task.

To prevent data contamination, retrieval results for a given target repository exclude its own package entry from the embedding index. When evaluating generation quality, only metadata and recipe chunks from other packages are used as retrieval candidates, ensuring that the context never contains the ground-truth. 

While this method and the graph-based strategy differ in how they define similarity, both operate over the E4S corpus and aim to enhance recipe generation. The embedding-based retrieval captures nuanced relationships that graphs may overlook, whereas graph-based affinity measures deterministic structural alignment. Our evaluation treats both as complementary retrieval strategies for understanding how different similarity signals influence recipe generation outcomes.

\noindent\textbf{No Reference.} 
To measure the ability of LLMs to generate valid Spack packages without reference information from an example Spack recipes, we support the capability to omit references in the generation workflow.

\subsection{Generation}

After the analysis and retrieval tools complete their respective stages, the agent assembles a prompt to the LLM using the collected data. The prompt includes a preamble describing the task of generating a Spack package file, along with relevant guidelines and common Spack heuristics that assist the model in interpreting the repository metadata. The preamble also provides concise, domain-specific instructions that reflect practical knowledge of Spack’s packaging conventions, such as typical directive usage, dependency patterns, and variant definitions. This context helps the model interpret repository information in a way that approximates how an experienced Spack contributor would reason about a new package.

\subsection{Evaluation Tools}
\label{subsec:eval-tool}
Each generated recipe is validated through a multi-stage pipeline that mirrors Spack's native build process. The system first verifies that the recipe can be parsed, then resolves its dependencies and configuration options to produce a concrete build plan, and finally attempts to compile and install the target software. After functional checks, the tool computes structural similarity metrics as described in Section~\ref{subsec:similarity-metrics}.

Testing package installs in HPC environments is challenging due to complex dependency structures and system-specific configurations. Each evaluation is therefore executed in an isolated Ubuntu 22.04 container with a clean build environment to prevent cross-contamination from pre-existing libraries or tools. Any dependencies may be pulled from a Spack \emph{buildcache}, a repository of pre-built binaries used to reduce overall install time.

\subsection{Error-Aware Repair Loop}
\label{subsec:repair-loop}

When evaluation fails, the agent initiates an iterative recovery phase to repair the generated recipe using feedback from the failed build.
For each unsuccessful iteration, the agent retains the erroneous recipe, the associated error message, any Spack references included in the previous context, and a condensed version of the original generation prompt. These elements form the input for the next model invocation, enabling the LLM to reason about the failure and propose a corrected recipe. The loop continues until the package builds successfully or a maximum of $k$ repair attempts is reached.
This process evaluates both the model’s capacity for self-correction and the utility of Spack’s diagnostic output. Empirically, it measures how structured error feedback can guide iterative refinement toward a valid, buildable package.

For further error information, we utilize the Spack provided \texttt{audit} command for static validation of packages, detecting issues such as missing download directives, nonexistent dependencies, or syntax errors. In our framework, audit checks are integrated into the error-aware repair loop as an additional feedback signal beyond build outcomes. When a build fails, the framework runs \texttt{spack audit} on the generated file and includes the resulting diagnostics in the next repair iteration. This enriches the feedback context enabling the model to interpret both compile-time failures and static configuration defects. The audit phase tests whether diagnostic feedback improves convergence toward a valid package.

%% file: metrics.tex
Determining correctness for package metadata is non-trivial, as there are many different ways to define the same package.
For instance, a user might define a Spack variant to map to a superset of compiler flags, while the agent might define individual package variants for each flag in the superset.
Both of these scenarios could be interpreted as correct depending on the context.
Furthermore, if a generated Spack recipe builds and installs a software but misses some important optimization flags (e.g. fast math), it may be only partially correct.
Since no apparent single metric fully captures these nuances, we employ several criteria, including

functionality, correctness, and dependency and variant matching.

\subsection{Functionality and Correctness Measures}

Spack package installation proceeds through a series of validation stages that collectively assess the functional correctness of a generated recipe. Understanding where failures occur provides insight into which aspects of a specification are incomplete or inconsistent. 
A generated recipe must (1) \emph{load} as valid Python so Spack can interpret its structure. It must then (2) resolve dependency and variant constraints into a consistent configuration, a process known as \emph{concretization}~\cite{gamblin_using_2022}, and finally it must (3) \emph{build and install} the target software successfully.

We consider loading, concretization, and installation success to be robust indicators of a model's ability to generate recipes.

Testing results can be useful in certain scenarios, but the definition of a test can vary widely among software project. For these reasons, we focus on Spack’s built-in validation stages, which provide standardized and reliable measures of correctness across all targets. 

The outcome of each validation stage provides insight into the quality of the generated package. A method may achieve a high rate of loaded recipes because it excels at writing Python, yet still fail during concretization if it cannot express Spack's constraint syntax or dependency logic. Even 
installation success alone is an insufficient metric.

Models can produce packages that load, concretize, and install correctly but omit essential details such as optional dependencies, conflicts, or variant definitions.

An example of this is shown in Figure~\ref{fig:spack-bad-example}, where the generated recipe completes all validation stages but lacks meaningful configuration metadata. Such specifications are incomplete within the Spack ecosystem, which relies on rich constraint information to support the customization of installations and tracking of build provenance. We point the reader to the Cgal package recipe\footnote{\href{https://github.com/spack/spack-packages/blob/1562c2ce3066ec4249a7b8dae844cccb0f7094a9/repos/spack\_repo/builtin/packages/cgal/package.py}{\url{https://github.com/spack/spack-packages/blob/develop/repos/spack\_repo/builtin/packages/cgal/package.py}}} that contains 182 lines of constraint, dependency, and flag specifications, illustrating the level of detail required for a complete recipe.

\begin{figure}[h]
  \centering
  \begin{lstlisting}[style=acmcode, language=Python]
class Cgal(CMakePackage):
    version('6.0.2', url='https://github.com/CGAL/cgal...')
  \end{lstlisting}
  \caption{An example Spack recipe that installs successfully, but omits essential details such as optional dependencies, debug/release flags, and version conflicts.}
  \label{fig:spack-bad-example}
  \Description{Python code showing a Spack package recipe with a defined tarball location.}
\end{figure}

\subsection{Similarity Measures}
\label{subsec:similarity-metrics}

While load, concretize, and install capture core functionality, they omit finer configuration nuance such as optional flags or build settings. Because Spack metadata can encode equivalent behavior in different forms, e.g., a single \texttt{release} variant that sets \texttt{-O3}/\texttt{-DNDEBUG} versus two variants (\texttt{optimization-level}, \texttt{debug}) that yield the same flags, exact parity is ill-defined, making general equivalence tests infeasible. We therefore complement functionality metrics with similarity measures. Generic code-similarity metrics like BLEU~\cite{papineni_bleu_2002} and CodeBLEU~\cite{ren_codebleu_2020} (a weighted combination of BLEU, weighted n-gram, AST, and data-flow signals) are informative for code but insensitive to package-specific semantics. Accordingly, we introduce \(S_v\) (variant similarity) and \(S_d\) (dependency similarity) tailored to Spack recipes, defined below.

\noindent\textbf{Variants.}\label{subsec:variant-score}
To quantify how faithfully a generated recipe preserves the configuration logic of the original, we define a variant similarity score \(S_v\), shown in Eq.~\ref{eq:variant-sim}. This metric measures the fraction of build arguments present in the model-generated recipe that also appear in the human-written recipe.

\vspace{3em}
\begin{equation}\label{eq:variant-sim}
    S_v =
    \dfrac{
        \left|\,\eqnmarkbox[RoyalBlue]{A}{A} \cap \eqnmarkbox[OliveGreen]{B}{B}\,\right|
    }{
        |\eqnmarkbox[RoyalBlue]{Am}{A}|
    }
\end{equation}
\annotate[yshift=1.5em,xshift=-10em]{above}{A}{Human written arguments}
\annotate[yshift=1.5em,xshift=0em]{above}{B}{Model generated arguments}

Formally, \(S_v\) is computed as the proportion of configuration argument keys shared between the two recipes, normalized by the number of arguments in the ground-truth recipe.
If the baseline defines no arguments (\(|A| = 0\)), the sample is excluded from scoring, since there is no configuration information to evaluate.
Configuration keys (e.g., \texttt{ENABLE\_MPI}) are statically extracted using the default Python AST parser, ensuring that the metric reflects structural similarity and remains independent of runtime conditions.

While configuration options are not strictly required in a Spack recipe, they convey valuable build-time options and represent an important part of the Spack ecosystem: providing fully parameterized package builds and installs. Because the ground-truth recipes have undergone Spack’s human review process, we regard their arguments as an established baseline of essential configurability. 

\noindent\textbf{Dependencies.}\label{subsec:dep-score}
We further propose a dependency similarity score $S_d$ to quantify how closely a generated recipe encodes the relationship between the package and its dependencies. Let \(D_A\) be the set of dependencies of the original package and \(D_B\) those of the generated package, after removing dependencies inherent to the package class.

Each dependency is represented as a tuple \((name, S, C, T)\) where \(name\) is the package name, \(S\) is the dependency specification, \(C\) is an optional condition, and \(T\) is a set of dependency types provided by the package (e.g., \texttt{build}, \texttt{link}, \texttt{run}).
For each dependency \(d_i \in D_A\), we find all dependencies in \(D_B\) sharing the same name and compute a composite match score for each candidate, defined in Eq.~\ref{eq:dep-match-score}.

\begin{equation}
\label{eq:dep-match-score}
M(d_i, d_j) = \alpha 
+ \beta \frac{|T_i \cap T_j|}{\max(|T_i|, 1)} 
+ \gamma \cdot \delta(S_i, S_j) 
+ \lambda \cdot \delta(C_i, C_j)
\end{equation}

Let \(\alpha+\beta+\gamma+\lambda=1\). \(\alpha\) rewards a name match; \(\beta\) rewards overlap in dependency types; \(\gamma\) exact spec equality; and \(\lambda\) condition equality.
During matching, each original dependency \(d_i \in D_A\) is paired with the most similar dependency in the generated recipe that shares its name. The mapping is non-exclusive: one generated dependency may cover multiple originals. This yields a coverage-oriented view, how well the generated recipe represents the originals, while the weighted score in Eq.~\ref{eq:dep-match-score} limits double counting by awarding high values only to strong matches. The overall dependency similarity is the mean best-match score:

\begin{equation}
S_d = \frac{1}{\max(|D_A|, 1)} 
\sum_{d_i \in D_A} 
\max_{d_j \in D_B,\, d_j.name = d_i.name} 
M(d_i, d_j)
\end{equation}

The metric ranges from \(0\) (no correspondence between dependency structures) to \(1\) (identical dependency configurations). It balances strict identity checks with partial credit for structurally similar dependencies, making it robust to minor metadata differences.
Unlike variant definitions, which can be represented as a flat set of strings, Spack dependencies are structured entities that include multiple dimensions.
A simple set overlap would fail to capture partial correspondence between dependencies that differ only slightly in configuration. For instance, a pure overlap metric would only compare package names, ignoring whether one package depends on \texttt{mpi} only for linking vs. linking and running.
We therefore define a weighted composite similarity score that awards partial credit for type and spec matches, yielding a more discerning and informative measure of structural similarity between dependencies.

See Appendix~\ref{sec:metrics-discussion} for additional discussion of these metrics and their advantages over generic code-similarity measures such as BLEU.

%% file: tasks.tex
To evaluate our framework, we construct a benchmark suite of Spack package generation tasks derived from the Extreme-scale Scientific Software Stack (E4S), described in ~\ref{sec:bg-spack}. E4S is a curated collection of high performance computing (HPC) software, each with an existing manually authored Spack recipe and a corresponding open-source repository. This combination provides a practical foundation for assessing automated build specification synthesis in HPC software environments.

\noindent\textbf{Source and Definition of Tasks.}
\label{subsec:e4s-tasks}
Our benchmark consists of 308 E4S packages that use the CMake build system. Each generation task corresponds to producing a Spack recipe for one target software repository, given its source tree and build configuration files. The selected repositories span core categories of HPC software, including numerical solvers (\texttt{Trilinos}), communication libraries, compiler frameworks (\texttt{LLVM}), and smaller utilities (\texttt{mozjpeg}). Collectively, they reflect the range of software components that define modern HPC ecosystems. As described in Section~\ref{sec:approaches}, an agentic system coordinates tools that extract repository metadata, retrieve relevant package examples, and synthesize a candidate recipe. The output is verified through a series of checks, with iterative repair applied when necessary. This process enables consistent, automated evaluation of model performance. 

\noindent\textbf{Rationale for Focusing on CMake.} 
While E4S includes projects that use a variety of build systems such as Autotools and GNU Make, we focus on CMake-based repositories. CMake provides a structured and widely adopted framework for describing build semantics, including dependency declarations with \texttt{find\_package}, configurable feature toggles, and explicit compiler requirements (e.g., \texttt{CMAKE\_CXX\_STANDARD}). These characteristics align well with Spack's declarative recipe model and make CMake repositories a natural starting point for this work. By focusing on CMake, we can develop and evaluate our generation approach on a subset of projects before extending it to other build systems with different conventions.

\noindent\textbf{Data Contamination Considerations.} 
Because the Spack recipes for E4S packages are publicly available, it is possible that some may have appeared in the training data of the language models we evaluate. To mitigate this risk and assess generation capabilities beyond existing Spack packages, we evaluate an additional set of 47 open-source repositories that we believe accurately represent the types of projects in the E4S stack and that are comparably difficult to package. These were selected from public GitHub projects containing CMake definitions and HPC-related keywords such as \texttt{solver}, \texttt{parallel}, and \texttt{numerical}. This extended set complements the E4S benchmark and helps isolate potential data contamination effects during evaluation.

%% file: setup.tex
Using the aggregated set of Spack packages, we evaluate the \spackagent{} approach on how well it can generate package recipes. This section details how we set up and run our experiments.

\noindent\textbf{Recipe Generation.}
\label{subsec:recipe-gen}
We evaluate recipe generation across the 308 E4S packages described in ~\ref{subsec:e4s-tasks}. For each experiment, a target repository is sampled at random from this set. Each generation task allows up to $k=5$ iterative attempts, enabling iterative refinement through the agentic workflow. All packages are installed without user-specified constraints to model the default behavior of Spack’s build specifications. Each generated recipe is evaluated using the tools described in Section~\ref{subsec:eval-tool}. When computing metrics, namely $S_d$, we use $\alpha=0.6$, $\beta=0.2$, $\gamma=0.1$, and $\lambda=0.1$. We found these to give the most intuitive scores for dependency matching.

The evaluation environment is designed to approximate the system configurations encountered by typical Spack users. We assess each generated recipe on an HPC system that provides a representative hardware and software environment for large-scale heterogeneous systems, reflecting the diverse environments that Spack must support. Each node contains a 64-core AMD EPYC~7713 CPU and four AMD~MI250X GPUs.

\noindent\textbf{\spackagent{} and Ablations.}
Our primary system follows the agentic workflow introduced in Section~\ref{sec:approaches}. The full configuration includes repository metadata extraction, retrieval of reference packages, static auditing of recipes, and iterative repair guided by build diagnostics. To assess the contribution of individual components, we perform ablation studies that disable one or more features. These include variations in the retrieval strategy, which may use structurally similar, embedding-based, random, random-within-build-system, or no reference packages. We disable components such as distillation, auditing, and the repair loop to measure their contribution to overall performance, and compare language models under these same conditions to isolate their respective impact on recipe validity and build success.

\noindent\textbf{Comparative Baseline (SWE-agent).} 
To contextualize results, we also evaluate \emph{SWE-agent} (introduced in Section \ref{sec:bg-llm}), adapting its automated repair workflow to the package generation setting. To adapt SWE-agent for our experiments, we reformulate each generation task into the issue-based format the agent expects. The instructions include references to relevant Spack commands that the agent can invoke to validate its edits. Once configured, SWE-agent executes its standard edit–verify–commit cycle to produce candidate patches. This configuration enables direct comparison between our domain-aware generation framework and a general-purpose software repair agent operating over structured build specifications.

\noindent\textbf{Inference.}
Open-weight models are executed using the vLLM inference engine~\cite{kwon_efficient_2023}, configured across four NVIDIA H100 GPUs. Closed-weight models are accessed through their respective API endpoints.

We evaluate both open- and closed-weight models, including a mix of reasoning and non-reasoning LLMs. 
The experiments compare GPT-5 \cite{openai_gpt5_2025}, GPT-4.1 \cite{openai_gpt4_1_2025}, Claude Sonnet 3.7 \cite{anthropic_claude_3_7_2025}, and Mistral Large 2.1 \cite{mistral_large_2_1_2024} on the generation tasks described earlier.

These models were selected to represent the current range of capabilities: GPT-5 and Claude Sonnet as advanced reasoning models, GPT-4.1 as a widely-adopted and documented baseline, and Mistral Large as a leading open-weight alternative. This diversity allows us to assess how reasoning capabilities and openness influence model performance in Spack package generation tasks.

%% file: results.tex
This section presents empirical evaluations of \spackagent{}, focusing on installation success, semantic correctness, and the effects of iterative reasoning, retrieval strategies, and feedback quality on Spack package generation. Detailed ablation results and supporting analysis are reported in Appendix~\ref{sec:ablation}, and illustrative case studies, including examples of successful repairs and persistent failure modes, are discussed in Appendix~\ref{sec:case-studies}.

\subsection{Extracting Repository Metadata}

\begin{rqbox}
\textbf{RQ1:} \textit{\RQone}
\end{rqbox}

To establish a baseline for automated Spack recipe generation, we first evaluate how effectively language models perform without iterative reasoning or retrieval assistance. A zero-shot configuration permits a single generation attempt without invoking any of the tools available to the agentic workflow, using only non-distilled repository metadata. This setting isolates the model's ability to infer valid build specifications without feedback or contextual grounding.

\noindent\textbf{Zero-shot baseline.}
When GPT-5 operates in this non-agentic mode only, 19.7\% of generated packages successfully install. Most failures occur during the concretization stage, indicating that the model often produces syntactically correct Python but incorrect dependencies or variant constraints. This behavior reflects Spack's structural complexity and confirms that build specification synthesis cannot be reliably performed through zero-shot generation.

\begin{figure}[t]
  \centering
  \includegraphics[width=\linewidth]{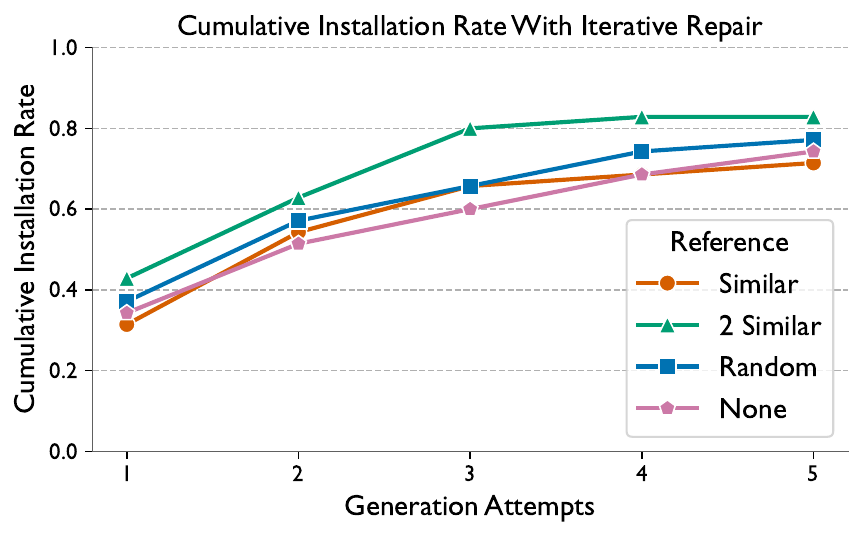}
  \caption{Cumulative installation success for GPT-5 over five repair attempts, grouped by reference type. All configurations improve with the agentic repair loop, with two similar references achieving the highest success rate (82.9\%).}
  \label{fig:gpt5-success-by-ref}
  
\end{figure}

\noindent\textbf{Effect of metadata distillation.}
By contrast, \spackagent{}'s full workflow (Section~\ref{sec:approaches}) integrates repository analysis, distillation, and retrieval to guide generation. Under the best-performing configuration (GPT-5 with two similar reference recipes and distilled CMake metadata), the system achieves 82.9\% installation success with up to five repair attempts (Figure~\ref{fig:gpt5-success-by-ref}). These results confirm that distilled, structured metadata significantly enhances model reasoning.

\input{tables/all}

\noindent\textbf{Cross-model summary.}
Table~\ref{tab:cross-model-combined} reports average performance metrics across all evaluated language models within five repair attempts ($k=5$). The evaluation set was balanced to ensure that each model was tested on comparable packages and feature coverage. GPT-5 achieved the highest installation and concretization success as well as the most accurate dependency structures, followed by GPT-4.1 and Claude Sonnet~3.7, with Mistral Large~2.1 performing lowest overall. These results show that reasoning-oriented, higher-capacity models are consistently more effective at generating valid and semantically complete Spack recipes.

For reference, approximately 76\% of the human-written E4S packages in our benchmark install successfully under the same environment. This rate serves as a practical reference point rather than a strict upper bound, as some upstream repositories or dependencies are no longer buildable and the human process differs fundamentally from automated generation.

Human-written and automatically generated recipes are not directly comparable. Human maintainers are not limited by a fixed number of attempts and continue refining a recipe until it builds correctly. They also benefit from community review, where feedback from experienced contributors enforces coding standards, cross-platform compatibility, and adherence to ecosystem conventions. Models, by contrast, operate with limited automated feedback and without this level of external validation. The human baseline therefore reflects collective maintenance effort and strict review rather than a bounded generation process.

\subsection{Iterative Repair and Feedback}
\label{sec:iterative-repair}

\begin{rqbox}
\textbf{RQ2:} \textit{\RQtwo}
\end{rqbox}

We next evaluate how the iterative repair process influences correctness of generated recipes. As described in Section~\ref{subsec:repair-loop}, \spackagent{} performs repeated repair attempts guided by Spack diagnostics until installation succeeds or a maximum number of steps is reached.

\noindent\textbf{Iterative improvement.}
Figure~\ref{fig:gpt5-success-by-ref} shows cumulative installation success for GPT-5 across repair attempts, segmented by the type of reference included in the context. When retrieval is enabled with two similar reference packages and distilled CMake metadata ,installation success rises rapidly reaching 82.9\% after five iterations. The success trajectory confirms that most recoverable errors are resolved within the first few cycles, and demonstrates that iterative reasoning over feedback is essential for achieving build success.

\begin{figure}[t]
  \centering
  \includegraphics[width=\linewidth]{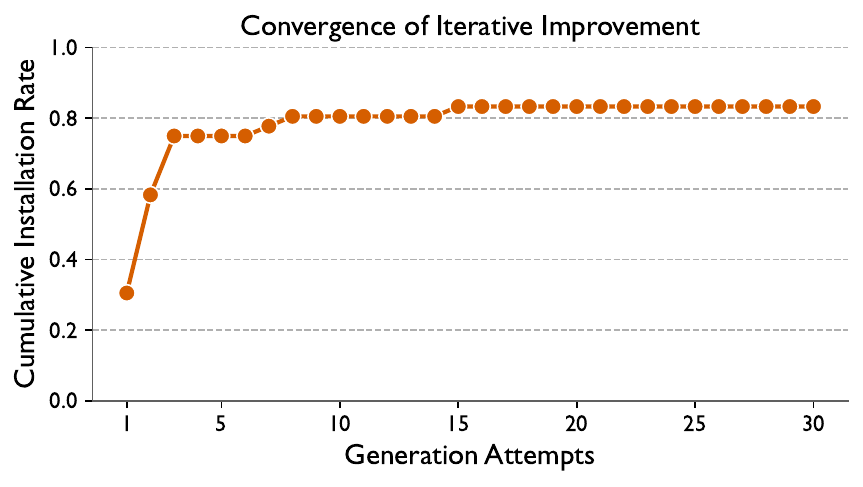}
  \caption{Cumulative installation success for GPT-5 with one similar reference across up to 30 repair attempts. Success rises rapidly through the first iterations, reaching 74\% by the fifth and plateauing near 83\% after 15 attempts.  }
  \label{fig:30attempts}
\end{figure}

\noindent\textbf{Convergence of improvement.}
To determine when the repair loop stops yielding benefits, we extended the maximum attempts to $k=30$. With a single similar reference and distilled repository metadata, installation success reached 74\% by the fifth attempt and plateaued at 83\% after roughly fifteen attempts (Figure~\ref{fig:30attempts}). Upon manual inspection, we found that two packages oscillated between distinct error states, alternately fixing one issue while reintroducing another. These results justify a cutoff between five and fifteen attempts, after which marginal improvement diminishes.

\noindent\textbf{Functional and variant validation.}
To verify that installation success corresponds to functional software and correct configuration logic, we conducted two validation studies. First, we manually tested twenty GPT-4.1 generated packages that installed successfully with a similar reference. Nineteen executed correctly on the AMD-based target system, while \texttt{Variorum} compiled but failed to configure architecture flags. 
Second, we examined thirty GPT-5 generated packages from the best-performing configuration (two similar references with distilled CMake). Five boolean variants per package were flipped to test conditional correctness and compared against human-authored recipes. LLM-generated packages installed successfully 74.2\% of the time versus 65.7\% for human-written baselines. Most remaining errors were due to missing dependencies that would have been detected by additional checks. Together, these results confirm that installation success corresponds to usable and semantically correct package recipes in most cases.

We categorize build failures into several types (e.g., syntax, concretization, dependency, and installation errors) to analyze where models most frequently struggle.
Definitions of these failure categories, along with their distributions across models and retrieval configurations, are presented in Appendix~\ref{sec:fail-tax}.

\subsection{Feedback Clarity and Refinement Velocity}

\begin{rqbox}
\textbf{RQ3:} \textit{\RQthree}
\end{rqbox}

\begin{figure}[t]
  \centering
  \includegraphics[width=\linewidth]{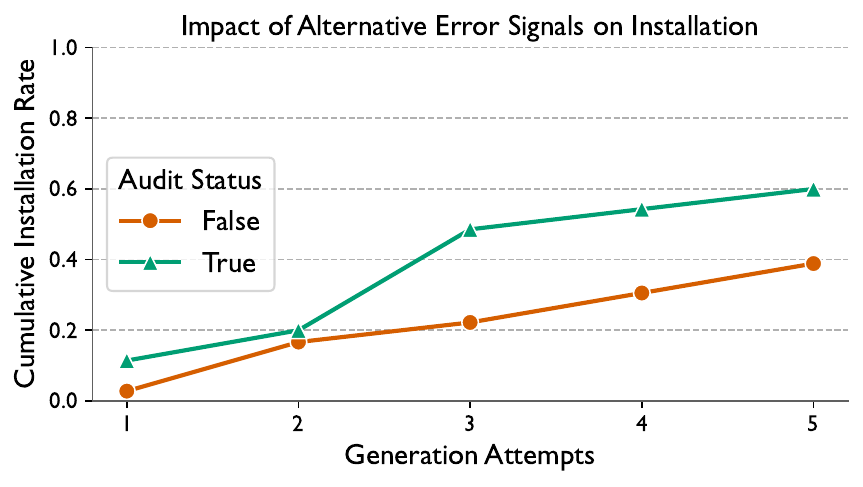}
  \caption{Installation success for GPT-4.1 with and without the \texttt{spack audit} feature. Including audit feedback improves success rates by providing clearer diagnostic signals, particularly benefiting lower-resource models within the repair loop.}
  \label{fig:audit}
\end{figure}

\noindent\textbf{Structured feedback through \texttt{spack audit}.}
As shown in Figure~\ref{fig:audit}, incorporating static validation of recipes (Section~\ref{subsec:repair-loop}) provides explicit feedback on missing fields, undefined dependencies, and syntactic issues that build logs often obscure. For a relatively low-performing configuration (GPT-4.1, distilled CMake, similar reference), installation success improved from 38.9\% to 60\% after five repair attempts when the audit step was enabled. The largest gains occurred between the second and third iterations, suggesting that early exposure to structured diagnostics accelerates convergence. This form of feedback particularly benefits smaller models, which depend more on explicit error signals than on implicit reasoning.

\subsection{Context Augmentation}

\begin{rqbox}
\textbf{RQ4:} \textit{\RQfour}
\end{rqbox}

As detailed in Section~\ref{subsec:retrieval-tool}, \spackagent{} supports multiple retrieval configurations that supply different types of contextual evidence, including structurally similar packages, random references, and embedding-based Spack-aware recipe chunks.

\begin{figure}[t]
  \centering
  \includegraphics[width=\linewidth]{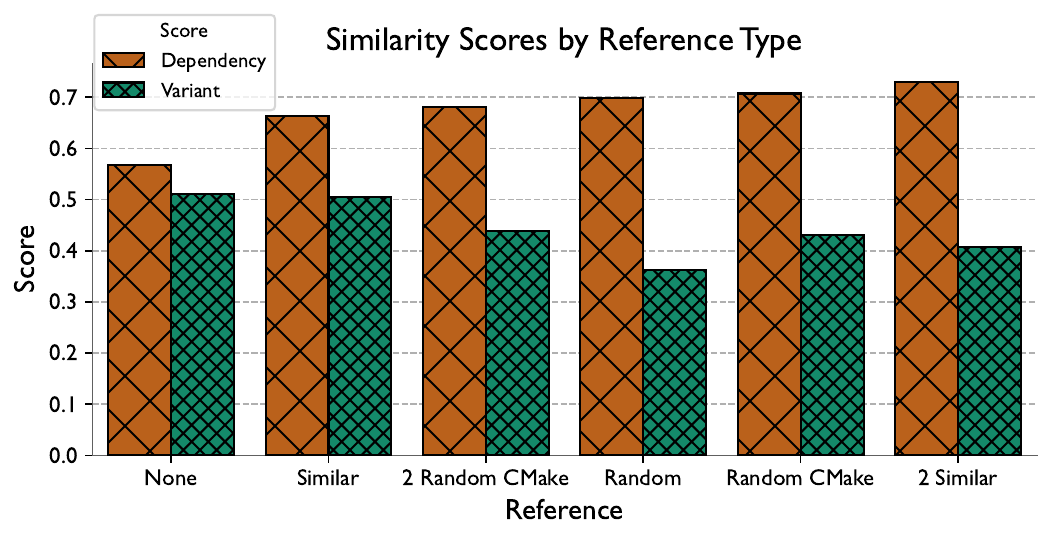}
  \caption{Dependency and variant similarity scores by reference type for GPT-5. Using two similar references yields the highest dependency similarity, while runs without references perform worst overall. Variant similarity shows no significant trend across reference types.}
  \label{fig:gpt5-scores-by-ref}
  \vspace{-1em}
\end{figure}

\noindent\textbf{Effect of reference type.}  
Figure~\ref{fig:gpt5-success-by-ref} shows cumulative installation success rates for GPT-5, segmented by reference type. Runs that included two structurally similar reference packages achieved the highest overall installation rate (82.9\%) confirming that structural alignment with known recipes provides the strongest contextual signal. Interestingly, even runs without any references show gradual improvement over successive repair attempts, indicating that GPT-5 possesses a baseline understanding of Spack's declarative structure, though it benefits substantially from explicit contextual grounding.

To further assess differences across configurations, Figure~\ref{fig:gpt5-scores-by-ref} compares dependency and variant similarity scores as defined in Section~\ref{subsec:similarity-metrics}. All reference-based configurations produced recipes with more complete and semantically accurate dependency graphs, a statistically significant improvement (\(p = 2.05\times10^{-6}\), \(d = 0.55\)). This suggests that retrieved context helps the model recover implicit relationships between libraries and build systems that are otherwise difficult to infer from repository metadata alone. Variant similarity showed no significant relationship across conditions, suggesting that modeling complex build-option logic remains difficult even with additional context. These trends persist across the full dataset, with a slightly smaller effect size (\(d = 0.49\)) for dependency similarity differences. Overall, the GPT-5 configuration with two similar references and the agentic repair loop yields the most complete and reliable Spack package representations.

\begin{figure}[t]
  \centering
  \includegraphics[width=\linewidth]{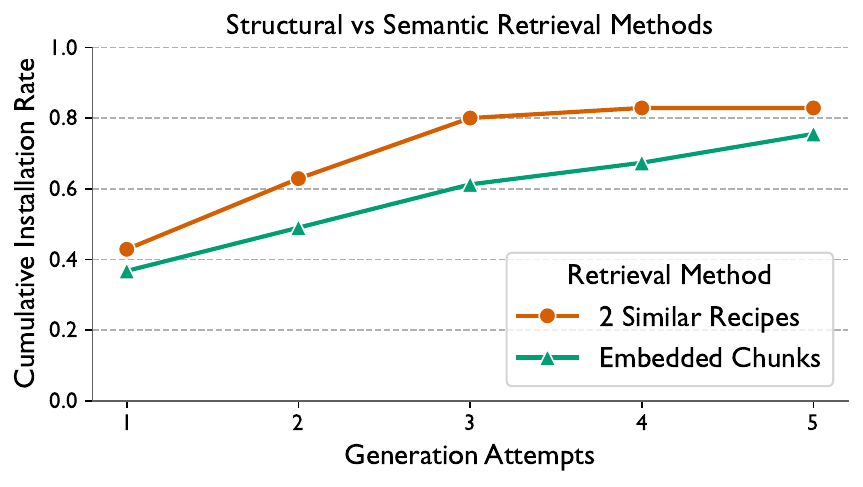}
  \caption{Cumulative installation success for GPT-5 comparing structural and semantic retrieval. The two-similar reference setup performs best, while the embedding-based RAG approach attains slightly lower performance.}
  \label{fig:gpt5-rag}
  \vspace{-1em}
\end{figure}

\noindent\textbf{Embedding-based retrieval.}
To evaluate semantic retrieval approaches, we also tested \spackagent{} using embedding-based Spack recipe chunks derived from the E4S knowledge graph. Using GPT-5, this configuration performed slightly below the graph-based setup but followed a nearly identical improvement curve (Figure~\ref{fig:gpt5-rag}). These findings demonstrate that both structural and semantic retrieval contribute to improved generation quality, and that combining them may further enhance future systems.

\noindent\textbf{Comparison with SWE-agent.}
To contextualize \spackagent{}'s domain-specific design, we compared it to the general-purpose SWE-agent framework (Section~\ref{sec:bg-llm}). Both systems were evaluated for a single generation attempt on a random sample of twenty packages using GPT-4.1. The raw, non-agentic baseline achieved a 5.9\% installation success rate, while SWE-agent reached 11.1\%. 
These results suggest comparable effectiveness within a single iteration. However, each SWE-agent invocation consumed an average of 725{,}829 tokens, compared to 36{,}481 for the raw baseline, roughly a twentyfold increase. This highlights that while SWE-agent can occasionally generate valid recipes, its generalized reasoning comes at a  financial cost, reinforcing the efficiency advantages of \spackagent{}'s targeted, domain-aware workflow for Spack package generation.

\begin{figure}[t]
  \centering
  \includegraphics[width=\linewidth]{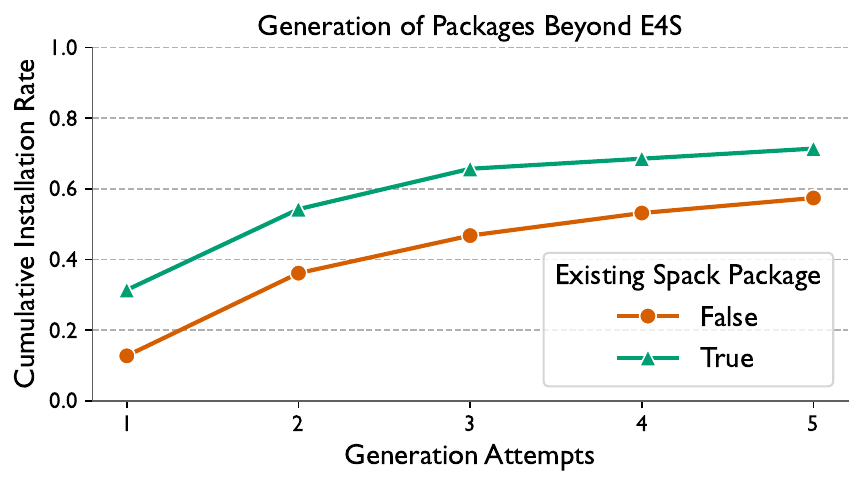}
  \caption{Installation success rates for GPT-5 on E4S (Spack) and non-Spack repositories. The agentic approach remains effective on unseen software, though success on non-Spack projects is $\approx$ 13\% lower, reflecting limited data familiarity.}
  \label{fig:gpt5-contamination}
  \vspace{-1em}
\end{figure}

\noindent\textbf{Generalization and data contamination.}
As discussed in Section~\ref{sec:tasks}, some E4S packages may exist in the public training data of large language models. To assess potential contamination, we evaluated \spackagent{} on open-source repositories not included in Spack, using installation success as the primary indicator of correctness. With GPT-5, one similar reference, and distillation enabled, the non-Spack repositories achieved installation rates about 13\% lower than E4S packages (Figure~\ref{fig:gpt5-contamination}). Despite this reduction, the agentic approach continued to produce a high proportion of valid installations, demonstrating strong generalization to previously unseen software and practical robustness under non-leaked conditions.

%% file: tables/all.tex
\begin{table}[t]
\centering
\caption{Average performance metrics across models within five repair attempts ($k=5$). Values represent mean success rates or similarity scores across all evaluated packages.}
\label{tab:cross-model-combined}
\setlength{\tabcolsep}{4pt}
\begin{tabular}{lrrr|r}
\toprule
\textbf{Model} & \textbf{Install} & \textbf{Dep. Score} & \textbf{Var. Score} & \textbf{$n$} \\
\midrule
GPT-5 & 0.78 & 0.66 & 0.46 & 285 \\
GPT-4.1 & 0.41 & 0.58 & 0.31 & 321 \\
Claude Sonnet~3.7 & 0.39 & 0.54 & 0.41 & 290 \\
Mistral Large~2.1 & 0.22 & 0.52 & 0.39 & 258 \\
\bottomrule
\end{tabular}
\end{table}

%% file: threats.tex
\noindent\textbf{Internal Validity.} LLM outputs are nondeterministic and can vary between runs. To reduce this variance, we run in a consistent containerized build environment and repeat experiments on subsets of the evaluation set to measure stability.

\noindent\textbf{Construct Validity.} Our evaluation relies on three indicators of correctness: successful installation, dependency similarity, and variant similarity. Installation success provides a concrete measure of functional correctness, while the similarity metrics quantify how well generated recipes reproduce human-authored configuration logic. These metrics have limitations. A recipe may install successfully but omit optional features, or achieve a high similarity score while expressing unnecessary directives. We mitigate these issues by assessing functional and structural measures in parallel and by manually validating a sampled subset of generated recipes.

\noindent\textbf{External Validity.} We evaluate on real-world HPC software from the E4S stack, not a simplified or synthetic benchmark. This ensures that results reflect practical packaging scenarios faced by Spack users. However, this focus limits generalization to other build systems and ecosystems. To assess generalization, we evaluate repositories without existing Spack packages, which achieved slightly lower but comparable performance.

%% file: related.tex
In this section we review related works on using LLMs to automate software configuration, compilation, and package management.

\noindent\textbf{Agents for Configuration and Compilation.}
CXXCrafter \cite{yu_cxxcrafter_2025} targets large C/C++ projects, identifying build systems, generating container manifests, and iteratively repairing errors from build logs; it reports strong build success on curated sets but does not emit declarative specifications or parameterized options. BuildBench~\cite{zhang_buildbench_2025} and CompileAgent~\cite{hu_compileagent_2025} introduce repo-level benchmarks where LLM agents search for build instructions, retrieve configuration details, and perform post-failure repairs across diverse open-source repos. Repo2Run~\cite{hu_repo2run_2025} constructs executable environments for Python repositories with tests by iteratively building containers and synthesizing instructions. Milliken et al.~\cite{milliken_beyond_2025} automate installation for Python-based repositories by extracting commands from documentation. Mondesire et al.~\cite{mondesire_automating_2025} propose a multi-agent framework for HPC system administration that generates build scripts and installs software via system package managers. 
Across these efforts, success is defined by ``it builds/runs." They do not extract or reason over structured build metadata, model the configuration space, or capture parameterized options; they are largely agnostic to how dependencies are resolved so long as installation succeeds. In contrast, we analyze build logic to reconstruct the configuration surface and dependency relationships, emitting declarative Spack recipes that describe the parameterized build space. 
This source-driven approach targets heterogeneous, multi-language HPC software and yields portable, reproducible specifications across compilers, architectures, and toolchains.

\noindent\textbf{LLM-Generated Package Metadata.}
Beyond build and environment automation, some studies investigate how language models can generate complete software packages or metadata files. PyGen~\cite{barua_pygen_2025} presents a system for the Python ecosystem that creates libraries from natural language prompts. The work demonstrates that LLMs can produce coherent package structures and complete metadata, highlighting the potential of language models to accelerate development workflows within mature ecosystems such as PyPI. 
While PyGen operates effectively within the standardized conventions of Python packaging, our domain is considerably more heterogeneous. The Spack ecosystem encompasses software written in multiple languages and built with diverse systems such as CMake, Autotools, and GNU Make, each exposing different configuration and dependency mechanisms. Our approach operates directly on this build metadata to generate declarative recipes that reflect a project’s dependency structure and configurable build options.

%% file: conclusion.tex
Building and maintaining packages for HPC software remains a persistent barrier to productivity in the field. In this work we introduced \spackagent{}, an agentic framework that automates the process of Spack package creation by emulating the workflow of expert packagers. \spackagent{} combines repository analysis, retrieval of relevant references, and iterative self-repair to synthesize Spack recipes. Across 308 CMake-based repositories, it improves installation success from 19.7\% in a zero-shot setting to nearly 83\% in its best configuration, while maintaining high semantic accuracy and strong token efficiency compared to general-purpose software engineering agents.

%% file: acknowledgements.tex
We thank Tamara Dahlgren for helpful discussions throughout this project.

This work was performed under the auspices of the U.S. Department of Energy (DOE) by Lawrence Livermore National Laboratory under Contract DE-AC52-07NA27344 (LLNL-CONF-2013107). This material is based upon work supported by the DOE Office of Science, Advanced Scientific Computing Research program through solicitation DE-FOA-0003264, "Advancements in Artificial Intelligence for Science."

%% file: appendix.tex
\section{Ablation Studies}
\label{sec:ablation}

To assess the contribution of individual components in the \spackagent{} workflow, we performed ablation experiments that varied retrieval strategies, repository metadata representations, and language models. The iterative repair loop was fixed at five attempts across all configurations, as this setting provided a practical balance between repair depth and runtime cost (see Section~\ref{sec:results}). The reported scores correspond to the successful attempt for each package. This section summarizes the impact of each factor on installation success and semantic correctness.

\noindent\textbf{Retrieval.} We evaluated multiple retrieval strategies to assess how the type of contextual information influences generation quality. Configurations included no retrieval, random package references, random references restricted by build system, and structurally similar reference packages. This variation measures how the relevance of retrieved context affects the model's ability to generate functional and semantically accurate recipes.

\noindent\textbf{Distillation.} Repository metadata was provided to the model either in its raw extracted form or as a distilled summary. The distilled representation compresses CMake metadata into a structured schema that highlights dependency declarations, feature flags, and compiler constraints. This comparison tests whether compact, structured metadata improves reasoning efficiency and contextual alignment relative to unfiltered input.

\noindent\textbf{Language Models.} We compared several large language models under identical experimental conditions to measure how model capability influences recipe generation. The evaluation includes GPT-5, GPT-4.1, Claude Sonnet~3.7, and Mistral Large~2.1, covering a range of reasoning depth and generative quality. These experiments isolate the effect of model capacity from the surrounding agent architecture.

\input{tables/ablation}

\subsection{Ablation Results}
\label{sec:ablation-res}

Table~\ref{tab:ablation_by_model} reports mean outcomes across language models and retrieval configurations, including the fraction of successful load, concretization, and installation stages, averaged semantic similarity scores for variants ($S_v$) and dependencies ($S_d$), and the mean number of repair attempts required for a successful installation. Each configuration varies by the presence of retrieval, the number and type of references, and whether repository metadata were provided in raw or distilled form.

Across all models, retrieval and metadata distillation generally improve installation success and semantic similarity, though the magnitude of these effects varies with model capacity. \spackagent{} benefits most from retrieval when using one or two similar references, achieving installation rates above~80\% with GPT-5. For GPT-4.1 and Claude~3.7, similar-reference retrieval modestly increases concretization success but yields mixed installation outcomes, suggesting that smaller models are more sensitive to the relevance of retrieved examples. By contrast, random references sometimes degraded performance, producing higher constraint and dependency errors as discussed in Section~\ref{sec:fail-tax}. Mistral~2.1 shows limited gains and occasional regressions under retrieval, consistent with its lower baseline reasoning ability.

Metadata distillation has a stabilizing effect across nearly all models, often improving dependency similarity and reducing repair attempts without increasing token cost. Compact, structured representations appear to focus model reasoning and reduce errors in dependency identification. Finally, differences across models dominate overall outcomes: GPT-5 consistently achieves the highest success and semantic accuracy, followed by GPT-4.1, Claude~3.7, and Mistral~2.1. At the same time, the results also show that the iterative repair process improves performance across all models, even when absolute success rates differ, indicating that structured feedback benefits reasoning regardless of scale. Together, these findings reinforce \spackagent{}'s design principle of combining domain-specific retrieval, metadata distillation, and iterative refinement to achieve valid package generation outcomes.

\section{Failure Taxonomy}
\label{sec:fail-tax}

To complement the aggregate installation rates reported in Section~\ref{sec:ablation-res}, we categorize the distinct failure modes observed during recipe generation. This taxonomy summarizes the most common failures and their relative frequency across \spackagent{} configurations.

\subsection{Failure Categories}

\noindent\textbf{Web Failures.} These errors occur when a source archive or other resource cannot be fetched from the internet. Such failures are unrelated to recipe syntax or dependency logic, but they prevent successful builds and appear as transient network errors during compilation.

\noindent\textbf{Incorrect Constraints.} These failures arise during Spack's concretization phase, when dependency or variant conditions cannot be resolved into a consistent build plan. Typical causes include unknown or misspelled dependency names and conflicting conditional expressions. Because Spack's concretization errors often lack complete diagnostic information, the root cause is frequently ambiguous. These are among the most frequent and heterogeneous failure cases.

\noindent\textbf{Missing Dependencies.} These occur when a generated recipe omits a dependency required by the build system or by a transitive component. The error often manifests during configuration or compilation as a missing header or library path. Most of these issues are resolved in subsequent repair attempts due to the clear diagnostic signal provided by the error message.

\noindent\textbf{CMake Failures.} These failures occur when the repository's build system cannot be configured correctly. This may happen if the project lacks a proper CMake setup or if the generated recipe invokes CMake with incorrect options.

\noindent\textbf{Syntax Errors.} Syntax failures are raised when the generated recipe cannot be parsed by the Python interpreter. Examples include indentation issues, misplaced directives, or missing delimiters. These issues often indicate a model's inability to follow prompt instructions or to generate valid Python code.

\noindent\textbf{Compilation Errors.} This residual category aggregates build errors that cannot be classified into the previous groups. They include compiler errors and other issues that may reflect problems in the project repository rather than in the generated recipe.

\subsection{Failure Incidence Analysis}

\input{tables/error_models}
\input{tables/error_ref}

Tables~\ref{tab:error_by_model_fraction} and~\ref{tab:error_by_reference_fraction} summarize the relative frequency of each failure category across models and retrieval conditions. Overall, syntax and constraint errors were the most prevalent sources of failure, reflecting the challenge of producing syntactically valid and semantically consistent Spack recipes. Web and CMake-related failures were comparatively rare, indicating that external connectivity and repository-level configuration issues contributed little to overall error incidence.

Across models, GPT-5 showed the most balanced distribution of errors and the fewest constraint-related failures, suggesting stronger reasoning about dependency logic compared to other models. GPT-4.1 produced proportionally more concretization and syntax errors, consistent with its lower overall installation success reported in Section~\ref{sec:ablation-res}. Claude Sonnet~3.7 and Mistral Large~2.1 exhibited the highest rates of syntax-related failures, implying weaker adherence to Spack's Pythonic DSL structure. In contrast, GPT-5 more often failed at later stages, such as missing dependencies or compilation, indicating that its generated recipes were structurally valid but occasionally incomplete.

Across retrieval strategies, syntax errors remained the dominant failure type, accounting for roughly half of all observed errors in most configurations. Other categories varied only slightly, making it difficult to identify clear or consistent trends attributable to retrieval strategy alone. Runs with structurally similar references showed modest reductions in constraint-related failures, but these effects were minor relative to the overall prevalence of syntax errors. These results suggest that while retrieval configuration may influence specific error types, its impact is overshadowed by the broader challenge of maintaining syntactic validity during generation.

\section{Case Studies}
\label{sec:case-studies}
In addition to the aggregate results presented in Section~\ref{sec:results}, we include several qualitative case studies that show how \spackagent{} behaves in practice. These examples cover typical outcomes observed during package generation, including successful builds, iterative repairs, and failure cases.

\subsection{CabanaPD: Generation and Upstream Submission}
CabanaPD is a GPU-enabled peridynamics application written in C++ \cite{sam_reeve_ornlcabanapd_2025}. It builds on two E4S ecosystem projects, Kokkos and Cabana, to simulate powder-based manufacturing processes. Its combination of accelerator support and I/O features makes it representative of HPC packages available in Spack.

\spackagent{} analyzed the CabanaPD repository, distilled its CMake metadata, retrieved two  similar reference packages, and generated a Spack recipe. The initial version failed during installation because of a missing dependency, but \spackagent{} produced a functional package on its first repair attempt. GPT-5 was used as the language model.

We validated the generated recipe in our containerized evaluation environment. The GPU-enabled configuration was built and installed using:

\vspace{0.2em}
\begin{verbatim}
spack install cabana-pd +rocm +hdf5 amdgpu_target=gfx90a
\end{verbatim}
\vspace{0.2em}

After installation, we executed CabanaPD's test suite on the AMD GPUs in our environment. All tests completed successfully, confirming that \spackagent{} correctly expressed the build logic and dependencies required for a valid package.

We submitted the generated recipe upstream as a new Spack package, cabana-pd\footnote{\url{https://github.com/spack/spack-packages/pull/2290}}. CabanaPD did not previously exist in the Spack repository, making this submission a concrete example of how \spackagent{} can introduce new GPU-enabled scientific applications into the ecosystem with minimal human intervention. While our framework never performs automatic submissions, this case illustrates that the resulting packages are ready for community review with only minor edits.

\begin{figure*}[t]
  \centering
  \begin{minipage}{0.39\textwidth}
  \begin{lstlisting}[style=acmcode, language=Python]
class Fxdiv(CMakePackage):
    depends_on("c", type="build")
    depends_on("cxx", type="build")

    depends_on("cmake@3.5:", type="build")
    depends_on("python", type="build")

    def cmake_args(self):
        return [
            self.define("FXDIV_BUILD_TESTS", False),
            self.define("FXDIV_BUILD_BENCHMARKS", False),
        ]
  \end{lstlisting}
  \end{minipage}\hfill
  \begin{minipage}{0.56\textwidth}
  \begin{lstlisting}[style=acmcode, language=Python]
class Fxdiv(CMakePackage):
    variant("inline_assembly", default=False, description="Use inline assembly")
    variant("tests", default=False, description="Build tests")
    variant("benchmarks", default=False, description="Build benchmarks")

    depends_on("cmake@3.5:", type="build")
    depends_on("c", type="build")
    depends_on("cxx", type="build")

    def cmake_args(self):
        args = [
            self.define_from_variant("FXDIV_USE_INLINE_ASSEMBLY", "inline_assembly"),
            self.define_from_variant("FXDIV_BUILD_TESTS", "tests"),
            self.define_from_variant("FXDIV_BUILD_BENCHMARKS", "benchmarks"),
        ]
        return args
  \end{lstlisting}
  \end{minipage}
\caption{Comparison of two FXdiv package implementations. The recipe on the left was written by a Spack contributor, and the one on the right was generated by \spackagent{}. The generated recipe achieved a variant similarity score of 1.0 and a dependency similarity score of 0.75 due to a missing Python dependency. Despite this high structural agreement, the CrystalBLEU score was 0.39, reflecting low lexical overlap due to differences in formatting, naming, and directive ordering.}
  \label{fig:fxdiv-compared}
  \vspace{-1em}
\end{figure*}

\subsection{Iterative Refinement and Dependency Correction}

In this example, the target repository is the PEXSI electronic structure solver. The model used was GPT-4.1, and \spackagent{} was configured to distill CMake metadata and retrieve a random reference recipe. The first generated recipe failed during concretization with the following error:

\vspace{0.2em}
\begin{verbatim}
==> Error: pexsi is unsatisfiable
\end{verbatim}
\vspace{0.2em}

While mostly inscrutable, this message typically indicates that the constraints encoded in the recipe could not be resolved into a consistent build plan. On the next attempt, \spackagent{} refined the recipe by removing the explicit OpenMP dependency, which does not exist as a package in Spack and is instead handled through compiler options. These changes allowed the recipe to concretize but resulted in a failed installation due to an incompatibility with the version of SuperLU provided during compilation.

In the subsequent repair iteration, the agent adjusted the SuperLU dependency to match the version specified in the CMake metadata, resulting in a successful installation. The progression from an unsatisfiable constraint to a valid installation illustrates how iterative refinement enables the language model to align with Spack conventions.

\subsection{Persistent Failure After Deep Repair}

As noted in Section~\ref{sec:iterative-repair}, increasing the repair depth beyond five iterations (\(k > 5\)) yielded limited improvement. This example illustrates one such case, involving the SALT package, where \spackagent{} failed to converge even after thirty repair attempts.

The configuration used GPT-5 with distilled CMake metadata and a single similar reference recipe. The first attempt failed during concretization with an unsatisfiable set of constraints, and subsequent iterations failed during CMake configuration due to missing LLVM components. Over successive attempts, the model tried a range of fixes, including altering configuration options and searching the filesystem for an existing LLVM installation.

Inspection of the recipe available in the Spack repository\footnote{\href{https://github.com/spack/spack-packages/blob/ef3ca03dfe7dd2e0b6e7cfaa0efa0bc716156a3c/repos/spack_repo/builtin/packages/salt/package.py}{\url{https://github.com/spack/spack-packages/blob/develop/repos/spack_repo/builtin/packages/salt/package.py}}} reveals that SALT depends on a version of LLVM built with both \texttt{clang} and \texttt{flang} support. Both features were correctly identified in the extracted CMake metadata, but the model never configured them as variants of the LLVM dependency in the recipe. The reference example also lacked these variants, limiting the model's ability to infer the correct dependency pattern. This failure suggests that future systems may require more advanced retrieval or reasoning mechanisms to resolve dependency relationships that are expressed as build options.

\section{Discussion of Similarity Metrics}
\label{sec:metrics-discussion}

This section provides additional discussion of the variant (\(S_v\)) and dependency (\(S_d\)) similarity metrics introduced in Section~\ref{subsec:similarity-metrics}. These metrics were designed to capture the structural alignment between LLM-generated and human-written Spack recipes, focusing on alternative measures of success beyond installation.

\subsection{Comparison with Existing Code Similarity Metrics}

Unlike BLEU~\cite{papineni_bleu_2002} or CodeBLEU~\cite{ren_codebleu_2020}, which emphasize textual or syntactic similarity, \(S_v\) and \(S_d\) evaluate semantic alignment in package recipes. CrystalBLEU~\cite{eghbali_crystalbleu_2023} refines BLEU and CodeBLEU to better capture meaningful similarity in code. It removes trivial or boilerplate \(n\)-grams that appear frequently across a corpus before scoring, reducing the bias toward shared syntax and common patterns that can inflate traditional BLEU-based measures.

In Figure~\ref{fig:fxdiv-compared}, the human-written and \spackagent{}-generated recipes for FXdiv are shown side by side. The generated recipe achieves a variant similarity score of 1.0, as it correctly defines the \texttt{tests} and \texttt{benchmarks} build options. Its dependency similarity score is 0.75, reflecting the omission of a Python dependency present in the reference recipe. The CrystalBLEU score is 0.39, indicating low token-level overlap despite high structural agreement. Because CrystalBLEU measures surface similarity, differences in directive ordering, naming, and formatting reduce the score even when the underlying dependency and variant structure are largely equivalent. In the context of package specifications, where correctness depends on capturing build relationships rather than reproducing syntax, lexical overlap provides limited insight into functional equivalence.

The \(S_v\) and \(S_d\) metrics address these limitations by capturing structural and semantic fidelity rather than textual form. \(S_v\) evaluates how completely a generated recipe reproduces the user-exposed configuration options in the reference, while \(S_d\) measures correspondence among dependency names, types, and conditions, granting partial credit for closely related structures. Together, they provide a more accurate measure of functional alignment and better reflect how human maintainers assess the quality and completeness of Spack packages.

\begin{figure}[t]
  \centering
  \small
  \begin{lstlisting}[style=acmcode]
MATCH (p:Package)
WHERE p.name <> "cabana-pd"
  AND NOT toLower(p.name) CONTAINS toLower("cabana-pd")
  AND "cmake" IN p.build_systems
OPTIONAL MATCH (p)-[:DEPENDS_ON]->(d:Package)
OPTIONAL MATCH (p)-[:HAS_VARIANT]->(v:Variant)
WITH p,
     size(apoc.coll.intersection(
         ["cmake","cabana","nlohmann-json","googletest","cxx","c"],
         collect(d.name))) AS dep_score,
     size(apoc.coll.intersection(
         ["hdf5","silo","tests"],
         collect(v.name))) AS var_score
WITH p, dep_score, var_score,
     (0.6 * dep_score + 0.4 * var_score) AS total_score
ORDER BY total_score DESC
LIMIT 2
RETURN p.name AS match_name, p.recipe AS recipe, total_score
  \end{lstlisting}
    \caption{Neo4j query used to retrieve CMake-based packages related to cabana-pd. The query filters by build system, excludes the target by name, computes dependency and variant overlap scores, and returns the highest scoring matches.}
  \label{fig:cypher_query}
\end{figure}

\section{\spackagent{} Configuration}
\label{sec:config}

This section summarizes the configuration of \spackagent{} used in the ablation experiments, including the prompt and query employed for similar package retrieval from the Neo4j knowledge graph. For additional details and instructions on how to run \spackagent{}, see the replication package at \url{https://github.com/spack/spack-it}.

The prompt shown in Figure \ref{fig:prompt} is an abridged version of the template used to guide recipe synthesis in the workflow described in Section~\ref{sec:approaches}. It provides the language models with a structured instruction set describing the packaging task, heuristics that clarify Spack features, and guidance on how to interpret included reference information.

Similar retrieval was performed using the Neo4j query shown in Figure~\ref{fig:cypher_query}. The query implements the Package Affinity Score introduced in Section~\ref{subsec:retrieval-tool}, ranking candidate packages by the overlap of their declared dependencies and variants. Packages with CMake build systems most similar to the target repository are returned for use as contextual references during prompt construction. This retrieval step supplies the example packages consumed by the generation prompt in the ablation experiments.

\begin{figure*}[h]
  \centering
  \begin{lstlisting}[style=acmcode]
# OBJECTIVE
Given the metadata, output a Spack package recipe.

# GUIDELINES
- Response must be plain text. No syntax highlighting backticks.
- Use ONLY information explicitly present in the metadata. You MUST NOT invent versions, variants, or dependencies.
- Do not query online sources.

# HEURISTICS
- If a builder class (e.g., CudaPackage, ROCmPackage) is strongly implied by:
    - the build system class (cmake -> CMakePackage)
    - or feature hints like "python", you should include that builder class.
- Lean towards Spack best practices consistent with the reference package.
- Do not "copy" or assume behavior from the reference package if the metadata does not point to that truth.

# Naming packages
- Name the class using PascalCase directly from the Spack spec name:
    - Example: "bricks" -> class Bricks(...)

# Import rules
- The package will require builder classes. Use the provided metadata and other information to determine what to import.
    Example:
    - cmake -> from spack_repo.builtin.build_systems.cmake import CMakePackage
    - cuda -> from spack_repo.builtin.build_systems.cuda import CudaPackage
- The builder classes should be followed by this import:
    `from spack.package import *`

------

PACKAGE NAME: {{ pkg_name }}
BUILD SYSTEM: {{ build_sys }}
FEATURE HINTS: {{ features }}
BUILD SYSTEM METADATA: {{ cmake_distilled }}

VERSION INFORMATION:
I am providing the version information to the project. Add the version directive and other directives needed to fetch the software.
{{ version }}

I am providing the directory structure for this repo to provide insight into the relationship between files.
{{ tree }}

{% for name, ref in references.items() %}
  {% if name.startswith("similar") %}
the recipe was found to be one of the most "similar" to the target package, based on the supplied metadata
{{ ref.recipe }}
  {% elif name.startswith("random_buildsys") %}
this recipe was randomly selected. it will provide useful heuristics to you about spack packages.
{{ ref.recipe }}
  {% elif name.startswith("random") %}
this recipe was randomly selected. it will provide useful heuristics to you about spack packages.
{{ ref.recipe }}
  {% endif %}
{% endfor %}
  \end{lstlisting}
  \caption{Abridged version of the prompt template used to generate Spack recipes. It includes heuristics to assist the model in understanding Spack features and other advice for consuming included references.}
  \label{fig:prompt}
\end{figure*}

%% file: tables/ablation.tex
\begin{table*}[t]
\centering
\caption{Ablation results across models and retrieval configurations. Each row reports the mean fraction of successful
load, concretization, and installation stages for generated Spack recipes, along with averaged semantic similarity
scores for variants ($S_v$) and dependencies ($S_d$), and the mean number of repair attempts required for a successful installation.
Metadata indicates whether repository information was used in its distilled or raw form. Reference types indicate the
retrieval condition used to augment context during generation. Higher values across stages and similarity metrics correspond to improved recipe validity and semantic completeness,
while fewer attempts indicate faster convergence.}
\label{tab:ablation_by_model}
\setlength{\tabcolsep}{5pt}
\begin{tabular}{lllrrrrrr}
\toprule
\textbf{Model} & \textbf{Reference Type} & \textbf{Metadata} & \textbf{$n$} &
\textbf{Load} & \textbf{Concretize} & \textbf{Install} &
\textbf{$S_v$ / $S_d$} & \textbf{Attempts} \\
\midrule
\multicolumn{9}{l}{\textit{GPT-5}} \\
\midrule
 & None & Distilled & 35 & 1.00 & 0.86 & 0.74 & 0.46 / 0.64 & 2.12 \\
 & None & Raw & 37 & 1.00 & 0.92 & 0.78 & 0.56 / 0.51 & 1.72 \\
 & 1 Similar & Distilled & 35 & 1.00 & 0.91 & 0.71 & 0.47 / 0.65 & 1.92 \\
 & 1 Similar & Raw & 35 & 1.00 & 0.94 & 0.83 & 0.54 / 0.68 & 1.55 \\  %
 & 2 Similar & Distilled & 35 & 1.00 & 0.97 & 0.83 & 0.41 / 0.73 & 1.76 \\
 & 1 Random & Distilled & 35 & 1.00 & 0.86 & 0.77 & 0.36 / 0.70 & 1.96 \\
 & 1 Random CMake & Distilled & 37 & 1.00 & 1.00 & 0.78 & 0.43 / 0.71 & 1.66 \\
 & 2 Random CMake & Distilled & 36 & 1.00 & 0.94 & 0.78 & 0.44 / 0.68 & 1.64 \\
\midrule
\multicolumn{9}{l}{\textit{GPT-4.1}} \\
\midrule
 & None & Distilled & 38 & 0.97 & 0.55 & 0.45 & 0.26 / 0.60 & 2.06 \\
 & None & Raw & 38 & 1.00 & 0.84 & 0.53 & 0.21 / 0.24 & 2.05 \\
 & 1 Similar & Distilled & 36 & 1.00 & 0.67 & 0.39 & 0.29 / 0.62 & 3.14 \\
 & 1 Similar & Raw & 36 & 1.00 & 0.83 & 0.42 & 0.28 / 0.42 & 1.60 \\  %
 & 2 Similar & Distilled & 36 & 1.00 & 0.58 & 0.42 & 0.40 / 0.72 & 3.13 \\
 & 1 Random & Distilled & 64 & 0.98 & 0.56 & 0.38 & 0.34 / 0.67 & 2.50 \\
 & 1 Random CMake & Distilled & 36 & 0.97 & 0.61 & 0.44 & 0.31 / 0.67 & 2.56 \\
 & 2 Random CMake & Distilled & 37 & 1.00 & 0.60 & 0.27 & 0.38 / 0.64 & 2.30 \\
\midrule
\multicolumn{9}{l}{\textit{Claude Sonnet 3.7}} \\
\midrule
 & None & Distilled & 41 & 0.93 & 0.76 & 0.49 & 0.33 / 0.51 & 2.95 \\
 & None & Raw & 35 & 0.97 & 0.80 & 0.43 & 0.44 / 0.39 & 2.87 \\
 & 1 Similar & Distilled & 35 & 0.94 & 0.63 & 0.49 & 0.44 / 0.67 & 2.59 \\
 & 1 Similar & Raw & 35 & 0.83 & 0.51 & 0.31 & 0.32 / 0.39 & 3.00 \\  %
 & 2 Similar & Distilled & 36 & 0.89 & 0.56 & 0.33 & 0.46 / 0.51 & 3.17 \\
 & 1 Random & Distilled & 35 & 0.97 & 0.66 & 0.34 & 0.40 / 0.61 & 2.50 \\
 & 1 Random CMake & Distilled & 36 & 0.97 & 0.75 & 0.47 & 0.45 / 0.65 & 2.77 \\
 & 2 Random CMake & Distilled & 37 & 1.00 & 0.57 & 0.27 & 0.45 / 0.61 & 3.30 \\
\midrule
\multicolumn{9}{l}{\textit{Mistral Large 2.1}} \\
\midrule
 & None & Distilled & 33 & 0.79 & 0.46 & 0.30 & 0.35 / 0.64 & 3.10 \\
 & None & Raw & 36 & 0.81 & 0.42 & 0.22 & 0.55 / 0.38 & 2.38 \\
 & 1 Similar & Distilled & 33 & 0.88 & 0.42 & 0.24 & 0.33 / 0.48 & 3.38 \\
 & 1 Similar & Raw & 27 & 0.89 & 0.52 & 0.30 & 0.28 / 0.44 & 2.88 \\  %
 & 2 Similar & Distilled & 30 & 0.97 & 0.27 & 0.13 & 0.40 / 0.51 & 3.25 \\
 & 1 Random & Distilled & 35 & 0.94 & 0.49 & 0.26 & 0.41 / 0.60 & 3.44 \\
 & 1 Random CMake & Distilled & 29 & 0.90 & 0.28 & 0.14 & 0.37 / 0.46 & 2.50 \\
 & 2 Random CMake & Distilled & 35 & 0.89 & 0.49 & 0.17 & 0.40 / 0.62 & 3.17 \\
\bottomrule
\end{tabular}
\end{table*}

%% file: tables/error_models.tex
\begin{table*}[t]
\centering
\caption{Error incidence by model. Each value indicates the proportion of classified errors by type across all test cases for that model. Columns are ordered by error category.}
\label{tab:error_by_model_fraction}
\setlength{\tabcolsep}{6pt}
\begin{tabular}{lrrrrrrr}
\toprule
\textbf{Model} & \textbf{$n$} & \textbf{Web} & \textbf{Constraint} & \textbf{Missing Dep.} & \textbf{CMake} & \textbf{Syntax} & \textbf{Compilation} \\
\midrule
GPT-5 & 174 & 0.029 & 0.149 & 0.339 & 0.086 & 0.253 & 0.144 \\
GPT-4.1 & 185 & 0.016 & 0.308 & 0.216 & 0.059 & 0.303 & 0.097 \\
Claude Sonnet~3.7 & 213 & 0.014 & 0.113 & 0.085 & 0.033 & 0.709 & 0.047 \\
Mistral Large~2.1 & 121 & 0.017 & 0.273 & 0.141 & 0.025 & 0.537 & 0.008 \\
\bottomrule
\end{tabular}
\end{table*}

%% file: tables/error_ref.tex
\begin{table*}[t]
\centering
\caption{Error incidence by retrieval reference type. Each value indicates the proportion of classified errors by type across all test cases for that reference configuration. Columns are ordered by error category.}
\label{tab:error_by_reference_fraction}
\setlength{\tabcolsep}{6pt}
\begin{tabular}{lrrrrrrr}
\toprule
\textbf{Reference Type} & \textbf{$n$} & \textbf{Web} & \textbf{Constraint} & \textbf{Missing Dep.} & \textbf{CMake} & \textbf{Syntax} & \textbf{Compilation} \\
\midrule
1 Similar & 161 & 0.006 & 0.149 & 0.199 & 0.056 & 0.497 & 0.093 \\
2 Similar & 89 & 0.011 & 0.225 & 0.180 & 0.000 & 0.539 & 0.045 \\
1 Random & 103 & 0.029 & 0.223 & 0.243 & 0.058 & 0.340 & 0.107 \\
1 Random CMake & 85 & 0.035 & 0.271 & 0.153 & 0.059 & 0.424 & 0.059 \\
2 Random CMake & 67 & 0.000 & 0.134 & 0.224 & 0.060 & 0.478 & 0.104 \\
None & 188 & 0.027 & 0.218 & 0.175 & 0.064 & 0.452 & 0.064 \\
\bottomrule
\end{tabular}
\end{table*}

%% file: bib.bib
@inproceedings{spracklen_we_2025,
	title = {We {Have} a {Package} for {You}! {A} {Comprehensive} {Analysis} of {Package} {Hallucinations} by {Code} {Generating} {LLMs}},
	isbn = {978-1-939133-52-6},
	url = {https://www.usenix.org/conference/usenixsecurity25/presentation/spracklen},
	language = {en},
	urldate = {2025-09-25},
	author = {Spracklen, Joseph and Wijewickrama, Raveen and Sakib, A. H. M. Nazmus and Maiti, Anindya and Viswanath, Bimal},
	year = {2025},
	pages = {3687--3706},
}

@inproceedings{mondesire_automating_2025,
	address = {New York, NY, USA},
	series = {{PEARC} '25},
	title = {Automating {HPC} {Software} {Compilation}, {Deployment}, and {Error} {Resolution} through an {LLM}-based {Multi}-{Agent} {System}},
	isbn = {979-8-4007-1398-9},
	url = {https://dl.acm.org/doi/10.1145/3708035.3736023},
	doi = {10.1145/3708035.3736023},
	urldate = {2025-09-25},
	booktitle = {Practice and {Experience} in {Advanced} {Research} {Computing} 2025: {The} {Power} of {Collaboration}},
	publisher = {Association for Computing Machinery},
	author = {Mondesire, Sean and Nsiye, Emmanuel and Soykan, Bulent and Martin, Glenn},
	month = jul,
	year = {2025},
	pages = {1--8},
}

@inproceedings{gamblin_spack_2015,
	address = {New York, NY, USA},
	series = {{SC} '15},
	title = {The {Spack} package manager: bringing order to {HPC} software chaos},
	isbn = {978-1-4503-3723-6},
	shorttitle = {The {Spack} package manager},
	url = {https://dl.acm.org/doi/10.1145/2807591.2807623},
	doi = {10.1145/2807591.2807623},
	urldate = {2025-10-09},
	booktitle = {Proceedings of the {International} {Conference} for {High} {Performance} {Computing}, {Networking}, {Storage} and {Analysis}},
	publisher = {Association for Computing Machinery},
	author = {Gamblin, Todd and LeGendre, Matthew and Collette, Michael R. and Lee, Gregory L. and Moody, Adam and de Supinski, Bronis R. and Futral, Scott},
	month = nov,
	year = {2015},
	pages = {1--12},
}

@misc{noauthor_spack_2025,
	title = {Spack {Packaging} {Guide}},
	shorttitle = {Packaging {Guide}},
	url = {https://spack.readthedocs.io/en/latest/packaging_guide_creation.html},
	language = {en},
	urldate = {2025-10-09},
	month = oct,
	year = {2025},
}

@article{dubois_why_2003,
	title = {Why {Johnny} can't build [portable scientific software]},
	volume = {5},
	issn = {1558-366X},
	url = {https://ieeexplore.ieee.org/document/1225867},
	doi = {10.1109/MCISE.2003.1225867},
	number = {5},
	urldate = {2025-10-09},
	journal = {Computing in Science \& Engineering},
	author = {Dubois, P.F. and Epperly, T. and Kumfert, G.},
	month = sep,
	year = {2003},
	pages = {83--88},
}

@inproceedings{hoste_easybuild_2012,
	title = {{EasyBuild}: {Building} {Software} with {Ease}},
	shorttitle = {{EasyBuild}},
	url = {https://ieeexplore.ieee.org/abstract/document/6495863},
	doi = {10.1109/SC.Companion.2012.81},
	urldate = {2025-10-09},
	booktitle = {2012 {SC} {Companion}: {High} {Performance} {Computing}, {Networking} {Storage} and {Analysis}},
	author = {Hoste, Kenneth and Timmerman, Jens and Georges, Andy and De Weirdt, Stijn},
	month = nov,
	year = {2012},
	pages = {572--582},
}

@misc{kwon_efficient_2023,
	title = {Efficient {Memory} {Management} for {Large} {Language} {Model} {Serving} with {PagedAttention}},
	url = {http://arxiv.org/abs/2309.06180},
	doi = {10.48550/arXiv.2309.06180},
	urldate = {2025-10-09},
	publisher = {arXiv},
	author = {Kwon, Woosuk and Li, Zhuohan and Zhuang, Siyuan and Sheng, Ying and Zheng, Lianmin and Yu, Cody Hao and Gonzalez, Joseph E. and Zhang, Hao and Stoica, Ion},
	month = sep,
	year = {2023},
	note = {arXiv:2309.06180 [cs]},
}

@article{heroux_scalable_2024,
	title = {Scalable {Delivery} of {Scalable} {Libraries} and {Tools}: {How} {ECP} {Delivered} a {Software} {Ecosystem} for {Exascale} and {Beyond}},
	volume = {26},
	issn = {1558-366X},
	shorttitle = {Scalable {Delivery} of {Scalable} {Libraries} and {Tools}},
	url = {https://ieeexplore.ieee.org/abstract/document/10494039},
	doi = {10.1109/MCSE.2024.3384937},
	number = {1},
	urldate = {2025-10-10},
	journal = {Computing in Science \& Engineering},
	author = {Heroux, Michael A.},
	month = jan,
	year = {2024},
	pages = {9--18},
}

@misc{ren_codebleu_2020,
	title = {{CodeBLEU}: a {Method} for {Automatic} {Evaluation} of {Code} {Synthesis}},
	shorttitle = {{CodeBLEU}},
	url = {http://arxiv.org/abs/2009.10297},
	doi = {10.48550/arXiv.2009.10297},
	urldate = {2025-10-10},
	publisher = {arXiv},
	author = {Ren, Shuo and Guo, Daya and Lu, Shuai and Zhou, Long and Liu, Shujie and Tang, Duyu and Sundaresan, Neel and Zhou, Ming and Blanco, Ambrosio and Ma, Shuai},
	month = sep,
	year = {2020},
	note = {arXiv:2009.10297 [cs]},
}

@inproceedings{papineni_bleu_2002,
	address = {USA},
	series = {{ACL} '02},
	title = {{BLEU}: a method for automatic evaluation of machine translation},
	shorttitle = {{BLEU}},
	url = {https://dl.acm.org/doi/10.3115/1073083.1073135},
	doi = {10.3115/1073083.1073135},
	urldate = {2025-10-10},
	booktitle = {Proceedings of the 40th {Annual} {Meeting} on {Association} for {Computational} {Linguistics}},
	publisher = {Association for Computational Linguistics},
	author = {Papineni, Kishore and Roukos, Salim and Ward, Todd and Zhu, Wei-Jing},
	month = jul,
	year = {2002},
	pages = {311--318},
}

@inproceedings{lewis_retrieval-augmented_2020,
	address = {Red Hook, NY, USA},
	series = {{NIPS} '20},
	title = {Retrieval-augmented generation for knowledge-intensive {NLP} tasks},
	isbn = {978-1-7138-2954-6},
	urldate = {2025-10-15},
	booktitle = {Proceedings of the 34th {International} {Conference} on {Neural} {Information} {Processing} {Systems}},
	publisher = {Curran Associates Inc.},
	author = {Lewis, Patrick and Perez, Ethan and Piktus, Aleksandra and Petroni, Fabio and Karpukhin, Vladimir and Goyal, Naman and Küttler, Heinrich and Lewis, Mike and Yih, Wen-tau and Rocktäschel, Tim and Riedel, Sebastian and Kiela, Douwe},
	month = dec,
	year = {2020},
	pages = {9459--9474},
}

@misc{guo_lightrag_2025,
	title = {{LightRAG}: {Simple} and {Fast} {Retrieval}-{Augmented} {Generation}},
	shorttitle = {{LightRAG}},
	url = {http://arxiv.org/abs/2410.05779},
	doi = {10.48550/arXiv.2410.05779},
	urldate = {2025-10-15},
	publisher = {arXiv},
	author = {Guo, Zirui and Xia, Lianghao and Yu, Yanhua and Ao, Tu and Huang, Chao},
	month = apr,
	year = {2025},
	note = {arXiv:2410.05779 [cs]},
}

@misc{hu_repo2run_2025,
	title = {{Repo2Run}: {Automated} {Building} {Executable} {Environment} for {Code} {Repository} at {Scale}},
	shorttitle = {{Repo2Run}},
	url = {http://arxiv.org/abs/2502.13681},
	doi = {10.48550/arXiv.2502.13681},
	urldate = {2025-10-17},
	publisher = {arXiv},
	author = {Hu, Ruida and Peng, Chao and Wang, Xinchen and Xu, Junjielong and Gao, Cuiyun},
	month = may,
	year = {2025},
	note = {arXiv:2502.13681 [cs]},
}

@article{yu_cxxcrafter_2025,
	title = {{CXXCrafter}: {An} {LLM}-{Based} {Agent} for {Automated} {C}/{C}++ {Open} {Source} {Software} {Building}},
	volume = {2},
	issn = {2994-970X},
	shorttitle = {{CXXCrafter}},
	url = {https://dl.acm.org/doi/10.1145/3729386},
	doi = {10.1145/3729386},
	language = {en},
	number = {FSE},
	urldate = {2025-10-17},
	journal = {Proceedings of the ACM on Software Engineering},
	author = {Yu, Zhengmin and Zhang, Yuan and Wen, Ming and Nie, Yinan and Zhang, Wenhui and Yang, Min},
	month = jun,
	year = {2025},
	pages = {2618--2640},
}

@misc{barua_pygen_2025,
	title = {{PyGen}: {A} {Collaborative} {Human}-{AI} {Approach} to {Python} {Package} {Creation}},
	shorttitle = {{PyGen}},
	url = {http://arxiv.org/abs/2411.08932},
	doi = {10.48550/arXiv.2411.08932},
	urldate = {2025-10-17},
	publisher = {arXiv},
	author = {Barua, Saikat and Rahman, Mostafizur and Sadek, Md Jafor and Islam, Rafiul and Khaled, Shehenaz and Hossain, Md Shohrab},
	month = jun,
	year = {2025},
	note = {arXiv:2411.08932 [cs]},
}

@misc{zhang_buildbench_2025,
	title = {{BuildBench}: {Benchmarking} {LLM} {Agents} on {Compiling} {Real}-{World} {Open}-{Source} {Software}},
	shorttitle = {{BuildBench}},
	url = {http://arxiv.org/abs/2509.25248},
	doi = {10.48550/arXiv.2509.25248},
	urldate = {2025-10-17},
	publisher = {arXiv},
	author = {Zhang, Zehua and Bajaj, Ati Priya and Handa, Divij and Liu, Siyu and Raj, Arvind S. and Chen, Hongkai and Wang, Hulin and Liu, Yibo and Basque, Zion Leonahenahe and Nath, Souradip and Juneja, Vishal and Chapre, Nikhil and Shoshitaishvili, Yan and Doupé, Adam and Baral, Chitta and Wang, Ruoyu},
	month = sep,
	year = {2025},
	note = {arXiv:2509.25248 [cs]},
}

@misc{hu_compileagent_2025,
	title = {{CompileAgent}: {Automated} {Real}-{World} {Repo}-{Level} {Compilation} with {Tool}-{Integrated} {LLM}-based {Agent} {System}},
	shorttitle = {{CompileAgent}},
	url = {http://arxiv.org/abs/2505.04254},
	doi = {10.48550/arXiv.2505.04254},
	urldate = {2025-10-17},
	publisher = {arXiv},
	author = {Hu, Li and Chen, Guoqiang and Shang, Xiuwei and Cheng, Shaoyin and Wu, Benlong and Li, Gangyang and Zhu, Xu and Zhang, Weiming and Yu, Nenghai},
	month = may,
	year = {2025},
	note = {arXiv:2505.04254 [cs]},
}

@inproceedings{milliken_beyond_2025,
	address = {Montreal, QC, Canada},
	title = {Beyond pip {Install}: {Evaluating} {LLM} {Agents} for the {Automated} {Installation} of {Python} {Projects}},
	copyright = {https://doi.org/10.15223/policy-029},
	isbn = {979-8-3315-3510-0},
	shorttitle = {Beyond pip {Install}},
	url = {https://ieeexplore.ieee.org/document/10992488/},
	doi = {10.1109/SANER64311.2025.00009},
	urldate = {2025-10-17},
	booktitle = {2025 {IEEE} {International} {Conference} on {Software} {Analysis}, {Evolution} and {Reengineering} ({SANER})},
	publisher = {IEEE},
	author = {Milliken, Louis and Kang, Sungmin and Yoo, Shin},
	month = mar,
	year = {2025},
	pages = {1--11},
}

@misc{suresh_cornstack_2025,
	title = {{CoRNStack}: {High}-{Quality} {Contrastive} {Data} for {Better} {Code} {Retrieval} and {Reranking}},
	shorttitle = {{CoRNStack}},
	url = {http://arxiv.org/abs/2412.01007},
	doi = {10.48550/arXiv.2412.01007},
	urldate = {2025-10-20},
	publisher = {arXiv},
	author = {Suresh, Tarun and Reddy, Revanth Gangi and Xu, Yifei and Nussbaum, Zach and Mulyar, Andriy and Duderstadt, Brandon and Ji, Heng},
	month = mar,
	year = {2025},
	note = {arXiv:2412.01007 [cs]},
}

@misc{wang_coderag-bench_2025,
	title = {{CodeRAG}-{Bench}: {Can} {Retrieval} {Augment} {Code} {Generation}?},
	shorttitle = {{CodeRAG}-{Bench}},
	url = {http://arxiv.org/abs/2406.14497},
	doi = {10.48550/arXiv.2406.14497},
	urldate = {2025-10-20},
	publisher = {arXiv},
	author = {Wang, Zora Zhiruo and Asai, Akari and Yu, Xinyan Velocity and Xu, Frank F. and Xie, Yiqing and Neubig, Graham and Fried, Daniel},
	month = feb,
	year = {2025},
	note = {arXiv:2406.14497 [cs]},
}

@inproceedings{gamblin_using_2022,
	address = {Dallas, TX, USA},
	title = {Using {Answer} {Set} {Programming} for {HPC} {Dependency} {Solving}},
	copyright = {https://doi.org/10.15223/policy-029},
	isbn = {978-1-6654-5444-5},
	url = {https://ieeexplore.ieee.org/document/10046107/},
	doi = {10.1109/SC41404.2022.00040},
	urldate = {2025-10-21},
	booktitle = {{SC22}: {International} {Conference} for {High} {Performance} {Computing}, {Networking}, {Storage} and {Analysis}},
	publisher = {IEEE},
	author = {Gamblin, Todd and Culpo, Massimiliano and Becker, Gregory and Shudler, Sergei},
	month = nov,
	year = {2022},
	pages = {1--15},
}

@misc{yang_swe-agent_2024,
	title = {{SWE}-agent: {Agent}-{Computer} {Interfaces} {Enable} {Automated} {Software} {Engineering}},
	shorttitle = {{SWE}-agent},
	url = {http://arxiv.org/abs/2405.15793},
	doi = {10.48550/arXiv.2405.15793},
	urldate = {2025-10-21},
	publisher = {arXiv},
	author = {Yang, John and Jimenez, Carlos E. and Wettig, Alexander and Lieret, Kilian and Yao, Shunyu and Narasimhan, Karthik and Press, Ofir},
	month = nov,
	year = {2024},
	note = {arXiv:2405.15793 [cs]},
}

@misc{noauthor_extreme-scale_nodate,
	title = {The {Extreme}-scale {Scientific} {Software} {Stack} ({E4S}) {Project}},
	url = {https://e4s.io},
}

@misc{gu_what_2025,
	title = {What to {Retrieve} for {Effective} {Retrieval}-{Augmented} {Code} {Generation}? {An} {Empirical} {Study} and {Beyond}},
	shorttitle = {What to {Retrieve} for {Effective} {Retrieval}-{Augmented} {Code} {Generation}?},
	url = {http://arxiv.org/abs/2503.20589},
	doi = {10.48550/arXiv.2503.20589},
	urldate = {2025-10-22},
	publisher = {arXiv},
	author = {Gu, Wenchao and Chen, Juntao and Wang, Yanlin and Jiang, Tianyue and Li, Xingzhe and Liu, Mingwei and Liu, Xilin and Ma, Yuchi and Zheng, Zibin},
	month = mar,
	year = {2025},
	note = {arXiv:2503.20589 [cs]},
}

@inproceedings{mcilroy1968mass,
  title={Mass-produced software components},
  author={McIlroy, M Douglas and Buxton, J and Naur, Peter and Randell, Brian},
  booktitle={Proceedings of the 1st international conference on software engineering, Garmisch Pattenkirchen, Germany},
  pages={88--98},
  year={1968}
}

@article{lehman1980programs,
  title={Programs, life cycles, and laws of software evolution},
  author={Lehman, Meir M},
  journal={Proceedings of the IEEE},
  volume={68},
  number={9},
  pages={1060--1076},
  year={1980},
  publisher={IEEE}
}

@article{decan2019empirical,
  title={An empirical comparison of dependency network evolution in seven software packaging ecosystems},
  author={Decan, Alexandre and Mens, Tom and Grosjean, Philippe},
  journal={Empirical Software Engineering},
  volume={24},
  number={1},
  pages={381--416},
  year={2019},
  publisher={Springer}
}

@inproceedings{bogart2016break,
  title={How to break an API: cost negotiation and community values in three software ecosystems},
  author={Bogart, Christopher and K{\"a}stner, Christian and Herbsleb, James and Thung, Ferdian},
  booktitle={Proceedings of the 2016 24th ACM SIGSOFT International Symposium on Foundations of Software Engineering},
  pages={109--120},
  year={2016}
}

@inproceedings{dietrich2019dependency,
	author = {Dietrich, Jens and Pearce, David and Stringer, Jacob and Tahir, Amjed and Blincoe, Kelly},
	booktitle = {2019 IEEE/ACM 16th International Conference on Mining Software Repositories (MSR)},
	organization = {IEEE},
	pages = {349--359},
	title = {Dependency versioning in the wild},
	year = {2019}}

@article{gonzalez2009macro,
	author = {Gonzalez-Barahona, Jesus M and Robles, Gregorio and Michlmayr, Martin and Amor, Juan Jos{\'e} and German, Daniel M},
	journal = {Empirical Software Engineering},
	number = {3},
	pages = {262--285},
	publisher = {Springer},
	title = {Macro-level software evolution: a case study of a large software compilation},
	volume = {14},
	year = {2009}}

@misc{apt,
	author = {Jason Gunthorpe},
	howpublished = {Online},
	note = {https://www.debian.org/doc/manuals/apt-guide/},
	title = {{APT User's Guide}},
	year = {1998}}

@misc{cargo,
	howpublished = {Online},
	month = {March},
	note = {https://github.com/rust-lang/cargo},
	title = {{Cargo: The Rust package manager}},
	year = {2014}}

@misc{openai_gpt5_2025,
  author       = {OpenAI},
  title        = {GPT-5 System Card},
  year         = {2025},
  month        = aug,
  howpublished = {\url{https://cdn.openai.com/gpt-5-system-card.pdf}},
  note         = {Accessed: October 2025}
}

@misc{openai_gpt4_1_2025,
  author       = {OpenAI},
  title        = {Introducing GPT-4.1 in the API},
  year         = {2025},
  month        = apr,
  howpublished = {\url{https://openai.com/index/gpt-4-1/}},
  note         = {Accessed: October 2025}
}

@misc{mistral_large_2_1_2024,
  author       = {Mistral AI},
  title        = {Mistral Large 2.1 Model Card},
  year         = {2024},
  month        = nov,
  howpublished = {\url{https://docs.mistral.ai/models/mistral-large-2-1-24-11}},
  note         = {Accessed: October 2025}
}

@misc{anthropic_claude_3_7_2025,
  author       = {Anthropic},
  title        = {Claude 3.7 Sonnet System Card},
  year         = {2025},
  month        = feb,
  howpublished = {\url{https://assets.anthropic.com/m/785e231869ea8b3b/original/claude-3-7-sonnet-system-card.pdf}},
  note         = {Accessed: October 2025}
}

@misc{sam_reeve_ornlcabanapd_2025,
	title = {{ORNL}/{CabanaPD}: {Version} 0.4},
	copyright = {BSD 3-Clause "New" or "Revised" License},
	shorttitle = {{ORNL}/{CabanaPD}},
	url = {https://zenodo.org/doi/10.5281/zenodo.16997636},
	urldate = {2025-11-03},
	publisher = {Zenodo},
	author = {Sam Reeve and Pablo Seleson and Irán Shepherd and Isner, Austin and Kinan, Bezem and Patrick Diehl},
	month = aug,
	year = {2025},
	doi = {10.5281/ZENODO.16997636},
}

@inproceedings{eghbali_crystalbleu_2023,
	address = {New York, NY, USA},
	series = {{ASE} '22},
	title = {{CrystalBLEU}: {Precisely} and {Efficiently} {Measuring} the {Similarity} of {Code}},
	isbn = {978-1-4503-9475-8},
	shorttitle = {{CrystalBLEU}},
	url = {https://dl.acm.org/doi/10.1145/3551349.3556903},
	doi = {10.1145/3551349.3556903},
	urldate = {2025-11-05},
	booktitle = {Proceedings of the 37th {IEEE}/{ACM} {International} {Conference} on {Automated} {Software} {Engineering}},
	publisher = {Association for Computing Machinery},
	author = {Eghbali, Aryaz and Pradel, Michael},
	month = jan,
	year = {2023},
	pages = {1--12},
}
